\documentclass{article}%
\usepackage{amsfonts}
\usepackage{amsmath}
\usepackage{amssymb}
\usepackage{graphicx}%
\newcommand{\bpartial}{\mathop{\partial\kern -4pt\raisebox{.8pt}{$|$}}}
\newcommand{\bra}{\mathopen{[\kern-1.6pt[}}
\newcommand{\ket}{\mathclose{]\kern-1.5pt]}}
\newcommand{\bbra}{\mathopen{[\kern-2.2pt[\kern-2.3pt[}}
\newcommand{\bket}{\mathclose{]\kern-2.1pt]\kern-2.3pt]}}
\newcommand{\slg}{\mbox{\bfseries\slshape g}}

\newcommand{\slT}{\mbox{\bfseries\slshape T}}

\makeindex
\begin{document}

\title{ Freud's Identity of Differential Geometry, the Einstein-Hilbert Equations and
the Vexatious Problem of the Energy-Momentum Conservation in GR }
\author{{\footnotesize Eduardo A. Notte-Cuello}$^{1}${\footnotesize and Waldyr A.
Rodrigues Jr.}$^{2}$
\and $^{1}${\footnotesize Departamento de Matem\'{a}ticas, Universidad de La
Serena, }
\and {\footnotesize Av. Cisterna 1200, La Serena-Chile.}
\and $^{2}${\footnotesize Institute of Mathematics, \ Statistics and Scientific
Computation}
\and {\footnotesize IMECC-UNICAMP CP 6065 \ \ \ \ 13083-859 Campinas, SP, Brazil}\\{\footnotesize walrod@ime.unicamp.br \ \ enotte@userena.cl}}
\maketitle

\begin{abstract}
We reveal in a rigorous mathematical way using the theory of differential
forms, here viewed as sections of a Clifford bundle over a Lorentzian
manifold, the true meaning of Freud's identity of differential geometry
discovered in 1939 (as a generalization of\ results already obtained by
Einstein in 1916) and rediscovered in disguised forms by several people. We
show moreover that contrary to some claims in the literature there is not a
single (mathematical) inconsistency between Freud's identity (which is a
decomposition of the Einstein \textit{indexed} $3$-forms $\star\mathcal{G}%
^{\mathbf{a}}$ in two \textit{gauge dependent }objects) and the field
equations of General Relativity. However, as we show there is an obvious
inconsistency in the way that Freud's identity is usually applied in the
formulation of energy-momentum \textquotedblleft conservation
laws\textquotedblright\ in GR. In order for this paper to be useful for a
large class of readers (even those ones making a first contact with the theory
of differential forms) all calculations are done with all details (disclosing
some of the \textquotedblleft tricks of the trade\textquotedblright\ of the subject).

\end{abstract}
\tableofcontents

\section{Introduction}

In \cite{san0,san1,san2,santilli1, santilli2} (and in references therein)
several criticisms to General Relativity (GR), continuation of the ones
starting in \cite{santilli}, are made. It is argued there that GR is full of
inconsistencies, which moreover are claimed to be solved by \textquotedblleft
isogravitation theory\textquotedblright\ \cite{santilli0,santilli1,santilli22}%
. It is not our intention here to make a detailed review of the main ideas
appearing in the papers just quoted. One of our purposes here is to
\textit{prove} that a strong claim containing there, namely that the classical
Freud's identity \cite{freud} of differential geometry is incompatible with
the vacuum Einstein-Hilbert field equations of GR is \textit{wrong}. We take
the opportunity to recall that Freud's identity is directly related with
proposals for the formulation of an energy-momentum \textquotedblleft
conservation law\textquotedblright\footnote{The reason for the
\textquotedblleft\ \textquotedblright\ will become clear soon.} in GR
\cite{trautman}. This issue is indeed a serious and vexatious problem
\cite{sawu} since unfortunately, the proposals appearing in the literature are
full of misconceptions. Some of them we briefly discuss below.\footnote{For
more details see \cite{rodoliv2007}.}

A sample on the kind of the misconceptions associated to the interpretation of
Freud's identity (and which served as inspiration for preparing the present
paper) show up when we read\footnote{Please, consult \cite{santilli1} for
knowledge of the references mentioned in the quotation below.}, e.g., in
\cite{santilli1} :

{\small \textquotedblleft A few historical comments regarding the Freud
identity are in order. It has been popularly believed throughout the 20-th
century that the Riemannian geometry possesses only four identities (see,
e.g., Ref. [2h]). In reality, Freud [11b]\footnote{{\small Reference
\cite{freud} in the present paper.}} identified in 1939 a fourth identity
that, unfortunately, was not aligned with Einstein's doctrines and, as such,
the identity was ignored in virtually the entire literature on gravitation of
the 20-th century.}

{\small However, as repeatedly illustrated by scientific history, structural
problems simply do not disappear with their suppression, and actually grow in
time. In fact, the Freud identity did not escape Pauli who quoted it in a
footnote of his celebrated book of 1958 [2g]\footnote{{\small Reference
\cite{pauli} in the present paper.}}. Santilli became aware of the Freud
identity via an accurate reading of Pauli's book (including its important
footnotes) and assumed the Freud identity as the geometric foundation of the
gravitational studies presented in Ref. [7d]. Subsequently, in his capacity as
Editor in Chief of Algebras, Groups and Geometries, Santilli requested the
mathematician Hanno Rund, a known authority in Riemannian geometry [2i], to
inspect the Freud identity for the scope of ascertaining whether the said
identity was indeed a new identity. Rund kindly accepted Santilli's invitation
and released paper [11c] of 1991 (the last paper prior to his departure) in
which Rund confirmed indeed the character of Eqs. (3.10) as a genuine,
independent, fourth identity of the Riemannian geometry.}

{\small The Freud identity was also rediscovered by Yilmaz (see Ref.
[11d]\footnote{{\small Reference \cite{yi3} in the present paper.}} and papers
quoted therein) who used the identity for his own broadening of Einstein's
gravitation via an external stress-energy tensor that is essentially
equivalent to the source tensor with non-null trace of Ref. [11a], Eqs.3.6 ).
Despite these efforts, the presentation of the Freud identity to various
meetings and several personal mailings to colleagues in gravitation, the Freud
identity continues to be main vastly ignored to this day, with very rare
exceptions (the indication by colleagues of additional studies on the Freud
identify not quoted herein would be gratefully appreciated).\textquotedblright%
\smallskip}

The paper is organized as follows. In Section 2 we present some preliminaries
which fix our notations and serve the purpose to present the Einstein-Hilbert
equations of GR within the theory of differential forms, something that makes
transparent the nature of all the objects involved. In Section 3 we recall the
Einstein-Hilbert Lagrangian density $\mathcal{L}_{EH}$ and the first order
gravitational Lagrangian $\mathcal{L}_{g}$ and the resulting field equations.
In Section 4 we recall that the components of the \textquotedblleft%
$2$-forms\textquotedblright\ $\star\mathcal{S}_{\mu}$ \ (Eq.(\ref{21}))
differs by\footnote{See Eq.(\ref{det}) for the definition of $\sqrt
{\mathbf{-g}}$.} $\sqrt{-\mathbf{g}}$ from the components of the objects
$\mathfrak{U}_{\mu}^{\rho\sigma}$ (Eq.(\ref{n6})) defined by Freud. \ We then
explicitly show that there is no incompatibility between Einstein equations
and Freud's identity which is seen as a gauge dependent decomposition of the
Einstein $3$-forms $\star\mathcal{G}_{\mu}$ .\ In Section 5 we recall a real
tragic problem, namely that there are \textit{no} genuine conservation laws of
energy-momentum (and of course angular momentum) in GR. \ Now, the details of
the proofs in Section 4 are presented in details in the Appendix C, and as the
reader will see, is a arduous exercise on the algebra and calculus of the
theory of differential forms, mathematical objects which in this paper are
supposed to be sections of the Clifford bundle of differential forms over a
Lorentzian manifold. A summary of the main results of the Clifford bundle
formalism, containing the main identities necessary for the purposes of the
present paper is given in Appendix A.\footnote{A detailed presentation of the
subject may be found in \cite{rodoliv2007}.}

\section{Preliminaries}

A Lorentzian manifold structure is a triple $\mathbf{L=}(M,%
\slg
,\tau_{g})$ where $M$ is a \textit{real} $4$-dimensional manifold (which is
Hausdorff, paracompact, \ connected and noncompact, equipped with a Lorentzian
metric $%
\slg
\in\sec T_{0}^{2}M$ \ and oriented by $\tau_{g}\in\sec\bigwedge^{4}T^{\ast}%
M$.\medskip

A spacetime structure is a pentuple $\mathfrak{M}=(M,%
\slg
,D,\tau_{g},\uparrow)$ \ where $(M,%
\slg
,\tau_{g})$ is a Lorentzian manifold, $D$ is the Levi-Civita connection of $%
\slg
$ and $\uparrow$ is an equivalence in $\mathbf{L}$ defining time
orientation.\footnote{Details may be found, e.g., in \cite{rodoliv2007,sawu}%
}\medskip

It is well known that in Einstein's General Relativity Theory (GRT) each
gravitational field generated by a energy-momentum density $\mathbf{T}\in\sec
T_{0}^{2}M$ \ is modelled by an appropriate $\mathfrak{M}$
\cite{rodoliv2007,sawu} .\medskip

Once $\mathbf{T}\in\sec T_{0}^{2}M$ \ is given the field $%
\slg
$\textbf{ }is determined through Einstein equation\footnote{We use natural
units.},%
\begin{equation}
\mathbf{G=Ric}-\frac{1}{2}%
\slg
R=-\mathbf{T=}%
\slT
, \label{1n}%
\end{equation}
where $\mathbf{Ric}\in\sec T_{0}^{2}M$ $\ $is the \textit{Ricci }tensor, $R$
is the \textit{curvature scalar} and $\mathbf{G}\in\sec T_{0}^{2}M$ $\ $is the
\textit{Einstein }tensor.\medskip

Let $(\varphi,U)$ be a chart for $U\subset M$ with coordinates $\{x^{\mu}\}$.
\ A coordinate basis for $TU$ is $\{\partial_{\mu}=\frac{\partial}{\partial
x^{\mu}}\}$ and its \textit{dual basis} (i.e., a basis for $T^{\ast}U$) is $\{%
\mbox{\boldmath{$\gamma$}}%
^{\mu}=dx^{\mu}\}$. We introduce also an orthonormal basis $\{\mathbf{e}%
_{\mathbf{a}}\}$ for $TU$ and corresponding dual basis $\{%
\mbox{\boldmath{$\theta$}}%
^{\mathbf{a}}\}$ for $T^{\ast}U$.

We have%
\[
\mathbf{e}_{\mathbf{a}}=h_{\mathbf{a}}^{\mu}\partial_{\mu},\text{ \ }%
\mbox{\boldmath{$\theta$}}%
^{\mathbf{a}}=h_{\mu}^{\mathbf{a}}dx^{\mu},
\]%
\begin{equation}
h_{\mu}^{\mathbf{a}}h_{\mathbf{b}}^{\mu}=\delta_{\mathbf{b}}^{\mathbf{a}%
},\ \ h_{\nu}^{\mathbf{a}}h_{\mathbf{a}}^{\mu}=\delta_{\nu}^{\mu}.
\end{equation}

The metric field is expressed in those basis as,%
\begin{align}%
\slg
&  =g_{\mu_{\nu}}%
\mbox{\boldmath{$\gamma$}}%
^{\mu}\otimes%
\mbox{\boldmath{$\gamma$}}%
^{\nu},\nonumber\\%
\slg
&  =\eta_{\mathbf{ab}}%
\mbox{\boldmath{$\theta$}}%
^{\mathbf{a}}\otimes%
\mbox{\boldmath{$\theta$}}%
^{\mathbf{b}}, \label{2}%
\end{align}
where the matrix with entries $\eta_{\mathbf{ab}}$ is $\mathrm{diag}%
(1,-1,-1,-1)$.\medskip

Next we introduce a metric \texttt{g\thinspace}$\in\sec T_{2}^{0}M$ on the
cotangent bundle by:%
\begin{align}
\mathtt{g}  &  =g^{\mu\nu}\frac{\partial}{\partial x^{\mu}}\otimes
\frac{\partial}{\partial x^{v}},\nonumber\\
\mathtt{g}  &  =\eta^{\mathbf{ab}}\mathbf{e}_{\mathbf{a}}\otimes
\mathbf{e}_{\mathbf{b}}, \label{3n}%
\end{align}
where the matrix with entries $\eta^{\mathbf{ab}}$ is $\mathrm{diag}%
(1,-1,-1,-1)$ and $g^{\mu\nu}g_{\nu\lambda}=\delta_{\lambda}^{\mu}$.\medskip

We introduce also the \textit{reciprocal basis} of $\{\partial_{\mu}%
=\frac{\partial}{\partial x^{\mu}}\}$ and of $\{\mathbf{e}_{\mathbf{a}}\}$ as
being respectively the basis $\{\partial^{\mu}\}$ and $\{\mathbf{e}^{a}\}$ for
$TU$ such that
\begin{equation}%
\slg
(\partial_{\mu},\partial^{\nu})=\delta_{\mu}^{\nu}\text{, }%
\slg
(\mathbf{e}_{\mathbf{a}},\mathbf{e}^{\mathbf{b}})=\delta_{\mathbf{a}%
}^{\mathbf{b}}, \label{4n}%
\end{equation}
and the \textit{reciprocal basis} of $\{%
\mbox{\boldmath{$\gamma$}}%
^{\mu}=dx^{\mu}\}$ and $\{%
\mbox{\boldmath{$\theta$}}%
^{\mathbf{a}}\}$ as being respectively the basis $\{%
\mbox{\boldmath{$\gamma$}}%
_{\mu}\}$ and $\{%
\mbox{\boldmath{$\theta$}}%
_{\mathbf{a}}\}$ for $T^{\ast}U$ \ such that
\begin{equation}
\mathtt{g}(%
\mbox{\boldmath{$\gamma$}}%
_{\mu},%
\mbox{\boldmath{$\gamma$}}%
^{\nu})=\delta_{\mu}^{\nu}\text{, }\mathtt{g}(%
\mbox{\boldmath{$\theta$}}%
_{\mathbf{a}},%
\mbox{\boldmath{$\theta$}}%
^{\mathbf{b}})=\delta_{\mathbf{a}}^{\mathbf{b}}. \label{5n}%
\end{equation}

We now observe that $\mathbf{Ric},%
\slg
$ and $\mathbf{T}$ (and of course, also $%
\slT
$) can be considered $1$-form valued $1$-form fields, i.e., we can write%
\begin{align}
\mathbf{Ric}  &  =\mathcal{R}_{\mu}\otimes%
\mbox{\boldmath{$\gamma$}}%
^{\mu}=\mathcal{R}^{\mu}\otimes%
\mbox{\boldmath{$\gamma$}}%
_{\mu}=\mathcal{R}_{\mathbf{a}}\otimes%
\mbox{\boldmath{$\theta$}}%
^{\mathbf{a}}=\mathcal{R}^{\mathbf{a}}\otimes%
\mbox{\boldmath{$\theta$}}%
_{\mathbf{a}},\nonumber\\%
\slg
&  =%
\mbox{\boldmath{$\gamma$}}%
_{\mu}\otimes%
\mbox{\boldmath{$\gamma$}}%
^{\mu}=%
\mbox{\boldmath{$\gamma$}}%
^{\mu}\otimes%
\mbox{\boldmath{$\gamma$}}%
_{\mu}=%
\mbox{\boldmath{$\theta$}}%
_{\mathbf{a}}\otimes%
\mbox{\boldmath{$\theta$}}%
^{\mathbf{a}}=%
\mbox{\boldmath{$\theta$}}%
^{\mathbf{a}}\otimes%
\mbox{\boldmath{$\theta$}}%
_{\mathbf{a}},\nonumber\\%
\slT
&  =%
\slT
_{\mu}\otimes%
\mbox{\boldmath{$\gamma$}}%
^{\mu}=%
\slT
^{\mu}\otimes%
\mbox{\boldmath{$\gamma$}}%
_{\mu}=%
\slT
_{\mathbf{a}}\otimes%
\mbox{\boldmath{$\theta$}}%
^{\mathbf{a}}=%
\slT
^{\mathbf{a}}\otimes%
\mbox{\boldmath{$\theta$}}%
_{\mathbf{a}}, \label{6N}%
\end{align}
where the $\mathcal{R}_{\mu}=R_{\mu\nu}%
\mbox{\boldmath{$\gamma$}}%
^{\nu}\in\sec%
{\displaystyle\bigwedge\nolimits^{1}}
T^{\ast}M$ (or the $\mathcal{R}^{\mu}\in\sec%
{\displaystyle\bigwedge\nolimits^{1}}
T^{\ast}M$ or the $\mathcal{R}_{\mathbf{a}}\in\sec%
{\displaystyle\bigwedge\nolimits^{1}}
T^{\ast}M$ or the $\mathcal{R}^{\mathbf{a}}\in\sec%
{\displaystyle\bigwedge\nolimits^{1}}
T^{\ast}M$ ) are called the Ricci $1$-form fields and the$\
\slT
_{\mu}=%
\slT
_{\mu\nu}%
\mbox{\boldmath{$\gamma$}}%
^{\nu}\in\sec%
{\displaystyle\bigwedge\nolimits^{1}}
T^{\ast}M\ $(or the $%
\slT
^{\mu}\in\sec%
{\displaystyle\bigwedge\nolimits^{1}}
T^{\ast}M$ or the $%
\slT
_{\mathbf{a}}\in\sec%
{\displaystyle\bigwedge\nolimits^{1}}
T^{\ast}M$ or the $%
\slT
^{\mathbf{a}}\in\sec%
{\displaystyle\bigwedge\nolimits^{1}}
T^{\ast}M\ $)\ are called the (negative) energy-momentum 1-form
fields.\footnote{Keep in mind that $%
\slT
=-T_{\nu}^{\mu}$ $\partial_{\mu}\otimes%
\mbox{\boldmath{$\gamma$}}%
^{\nu}$ and $\mathbf{T}=T_{\nu}^{\mu}\partial_{\mu}\otimes%
\mbox{\boldmath{$\gamma$}}%
^{\nu}$.}\medskip

We also introduce the Einstein $1$-form fields ($\mathcal{G}_{\mu}%
,\mathcal{G}^{\mu},\mathcal{G}_{\mathbf{a}},\mathcal{G}^{\mathbf{a}}$) by
writing the Einstein tensor as
\begin{equation}
\mathbf{G}=\mathcal{G}_{\mu}\otimes%
\mbox{\boldmath{$\gamma$}}%
^{\mu}=\mathcal{G}^{\mu}\otimes%
\mbox{\boldmath{$\gamma$}}%
_{\mu}=\mathcal{G}_{\mathbf{a}}\otimes%
\mbox{\boldmath{$\theta$}}%
^{\mathbf{a}}=\mathcal{G}^{\mathbf{a}}\otimes%
\mbox{\boldmath{$\theta$}}%
_{\mathbf{a}}, \label{7N}%
\end{equation}
where, e.g.,
\begin{align}
\mathcal{G}_{\mu}  &  =G_{\mu\nu}%
\mbox{\boldmath{$\gamma$}}%
^{\nu}\text{, }\mathcal{G}_{\mathbf{a}}=G_{\mathbf{ab}}%
\mbox{\boldmath{$\theta$}}%
^{\mathbf{b}},\text{ etc...}\label{8n}\\
G_{\mu\nu}  &  =R_{\mu\nu}-\frac{1}{2}g_{\mu\nu}R.
\end{align}

We now write Einstein equation (Eq.(\ref{1n})) as a set of equations for the
Einstein $1$-form fields, i.e.%
\begin{equation}
\mathcal{G}_{\mu}=%
\slT
_{\mu}\text{ or }\mathcal{G}_{\mathbf{a}}=%
\slT
_{\mathbf{a}}. \label{9N}%
\end{equation}
\medskip

We denoted by $\star$ the Hodge dual operator and write the dual of
Eq.(\ref{9N}) as%

\begin{equation}
\star\mathcal{G}_{\mu}=\star%
\slT
_{\mu}\text{ or }\star\mathcal{G}_{\mathbf{a}}=\star%
\slT
_{\mathbf{a}}. \label{10n}%
\end{equation}

\section{Gravitational Lagragian Densities}

As it is well known the Einstein-Hilbert Lagrangian density is%
\begin{equation}
\mathfrak{L}_{EH}=\frac{1}{2}\star R=\frac{1}{2}R\tau_{%
\slg
}. \label{11n}%
\end{equation}
\medskip\ \ \ 

We can easily verify that $\mathfrak{L}_{EH}$ can be written, e.g., as:%
\begin{equation}
\mathfrak{L}_{EH}=\frac{1}{2}\mathcal{R}_{\mathbf{cd}}\wedge\star(%
\mbox{\boldmath{$\theta$}}%
^{\mathbf{c}}\wedge%
\mbox{\boldmath{$\theta$}}%
^{\mathbf{d}}) \label{12new}%
\end{equation}
where
\begin{equation}
\mathcal{R}_{\mathbf{d}}^{\mathbf{c}}=d\omega_{\mathbf{d}}^{\mathbf{c}}%
+\omega_{\mathbf{k}}^{\mathbf{c}}\wedge\omega_{\mathbf{d}}^{\mathbf{k}}\in\sec%
{\displaystyle\bigwedge\nolimits^{2}}
T^{\ast}M \label{13new}%
\end{equation}
are the \textit{curvature} $2$-form fields with each one of the $\mathcal{R}%
_{\mathbf{d}}^{\mathbf{c}}$ being given by
\begin{equation}
\mathcal{R}_{\mathbf{d}}^{\mathbf{c}}=\frac{1}{2}R_{\mathbf{d\hspace
{0.02cm}\hspace{0.02cm}kl}}^{\hspace{0.02cm}\hspace{0.02cm}\,\mathbf{c}}%
\mbox{\boldmath{$\theta$}}%
^{\mathbf{k}}\wedge%
\mbox{\boldmath{$\theta$}}%
^{\mathbf{l}}, \label{14new}%
\end{equation}
where $R_{\mathbf{d\hspace{0.02cm}\hspace{0.02cm}kl}}^{\hspace{0.02cm}%
\hspace{0.02cm}\mathbf{c}}$ are the components of the Riemann tensor in the
orthogonal basis and where the\ $\omega_{\mathbf{d}}^{\mathbf{c}}%
:=\omega_{\mathbf{ad}}^{\mathbf{c}}%
\mbox{\boldmath{$\theta$}}%
^{\mathbf{a}}$ are the the connection $1$-forms in the gauge defined by the
orthonormal bases $\{\mathbf{e}_{\mathbf{a}}\}$ and $\{%
\mbox{\boldmath{$\theta$}}%
^{\mathbf{a}}\}$, i.e., $D_{\mathbf{e}_{\mathbf{a}}}%
\mbox{\boldmath{$\theta$}}%
^{\mathbf{b}}=-\omega_{\mathbf{ac}}^{\mathbf{b}}%
\mbox{\boldmath{$\theta$}}%
^{\mathbf{c}}$.

We recall that this paper the components of the Ricci tensor are defined
according to the following convention \cite{choquet,rindler}
\begin{equation}
\mathbf{Ric=}R_{\mathbf{d\hspace{0.02cm}\hspace{0.02cm}ka}}^{\hspace
{0.02cm}\hspace{0.02cm}\,\mathbf{a}}%
\mbox{\boldmath{$\theta$}}%
^{\mathbf{d}}\otimes%
\mbox{\boldmath{$\theta$}}%
^{\mathbf{k}}. \label{ricci}%
\end{equation}
\medskip\ 

We recall moreover that \ Eq.(\ref{13new}) is called Cartan's second structure
equation (valid for an arbitrary connection). Cartan's first structure
equations reads (in an orthonormal basis) for a torsion free connection, which
is the case of a Lorentzian spacetime
\begin{equation}
d%
\mbox{\boldmath{$\theta$}}%
^{\mathbf{a}}=-\omega_{\mathbf{b}}^{\mathbf{a}}\wedge%
\mbox{\boldmath{$\theta$}}%
^{\mathbf{b}}. \label{cartan1}%
\end{equation}

Also, it is not well known \textit{as it should be} that $\mathfrak{L}_{EH}$
can be written as\footnote{Details may be found in \cite{rodoliv2007}.}:%
\begin{equation}
\mathfrak{L}_{EH}=\mathfrak{L}_{g}-d(%
\mbox{\boldmath{$\theta$}}%
^{\mathbf{a}}\wedge\star d%
\mbox{\boldmath{$\theta$}}%
^{\mathbf{a}}) \label{15 new}%
\end{equation}
with\footnote{An equivalent expression for $\mathcal{L}_{g}(%
\mbox{\boldmath{$\theta$}}%
^{\mathbf{a}},d%
\mbox{\boldmath{$\theta$}}%
^{\mathbf{a}})$ is given in \cite{wallner}. However the formula there does not
disclose that $\mathcal{L}_{g}$ contains a Yangs-Mill \ term, a gauge fixing
term and an auto interaction term (in the form of interaction of the
vorticities of the fields $%
\mbox{\boldmath{$\theta$}}%
^{\mathbf{a}}$), something that suggests according to us, a more realistic
interpretation of Einstein's gravitational theory, i.e., as a theory of
physical fields in the Faraday sense living and interacting with all matter
fields in Minkowski spacetime \cite{notte}.}%
\begin{equation}
\mathfrak{L}_{g}=-\frac{1}{2}d%
\mbox{\boldmath{$\theta$}}%
^{\mathbf{a}}\wedge\star d%
\mbox{\boldmath{$\theta$}}%
_{\mathbf{a}}+\frac{1}{2}\delta%
\mbox{\boldmath{$\theta$}}%
^{\mathbf{a}}\wedge\star\delta%
\mbox{\boldmath{$\theta$}}%
_{\mathbf{a}}+\frac{1}{4}\left(  d%
\mbox{\boldmath{$\theta$}}%
^{\mathbf{a}}\wedge%
\mbox{\boldmath{$\theta$}}%
_{\mathbf{a}}\right)  \wedge\star\left(  d%
\mbox{\boldmath{$\theta$}}%
^{\mathbf{b}}\wedge%
\mbox{\boldmath{$\theta$}}%
_{\mathbf{b}}\right)  , \label{16new}%
\end{equation}
where $\delta$ is the Hodge coderivative operator.

Then the total Lagragian for the gravitational plus matter field can be
written as%
\begin{equation}
\mathfrak{L=L}_{g}+\mathfrak{L}_{m}, \label{17}%
\end{equation}
where due the principle of minimal coupling $\mathfrak{L}_{m}$ depends on the
matter fields (represented by some differential forms\footnote{We emphasize
that the present formalism is applicable even to spinor fields, which as
proved in \cite{moro, rodoliv2007} can safely be represented by appropriate
classes of \textit{non} \textit{homogeneous} differential forms.}) and the $%
\mbox{\boldmath{$\theta$}}%
^{\mathbf{a}}$ (due to the use of the Hodge dual in the writing of
$\mathfrak{L}_{m}$).\medskip

The variational principle%
\begin{equation}%
\mbox{\boldmath{$\delta$}}%
{\displaystyle\int}
(\mathfrak{L}_{EH}+\mathfrak{L}_{m})=0, \label{v1}%
\end{equation}
or
\begin{equation}%
\mbox{\boldmath{$\delta$}}%
{\displaystyle\int}
(\mathfrak{L}_{g}+\mathfrak{L}_{m})=0, \label{v2}%
\end{equation}
then must give with the usual hypothesis that the boundary terms are null the
\textit{same} equations of motion. From Eq.(\ref{v1}) we get supposing that
$\mathfrak{L}_{m}$ does not depend explicitly on the $d%
\mbox{\boldmath{$\theta$}}%
_{\mathbf{a}}$ (principle of minimal coupling) that
\begin{align}
&
{\displaystyle\int}
\mbox{\boldmath{$\delta$}}%
(\mathfrak{L}_{EH}+\mathfrak{L}_{m})\nonumber\\
&  =%
{\displaystyle\int}
(%
\mbox{\boldmath{$\delta$}}%
\mathfrak{L}_{EH}+%
\mbox{\boldmath{$\delta$}}%
\mbox{\boldmath{$\theta$}}%
^{\mathbf{a}}\wedge\frac{\partial\mathfrak{L}_{m}}{\partial%
\mbox{\boldmath{$\theta$}}%
_{\mathbf{a}}}). \label{v4}%
\end{align}
The result of this variation is (see details in the Appendix B):%

\begin{align}
\int%
\mbox{\boldmath{$\delta$}}%
(\mathfrak{L}_{EH}+\mathfrak{L}_{m})  &  =\int%
\mbox{\boldmath{$\delta$}}%
\mbox{\boldmath{$\theta$}}%
^{\mathbf{a}}\wedge(\frac{%
\mbox{\boldmath{$\delta$}}%
\mathfrak{L}_{EH}}{%
\mbox{\boldmath{$\delta$}}%
\mbox{\boldmath{$\theta$}}%
^{\mathbf{a}}}+\frac{\partial\mathfrak{L}_{m}}{\partial%
\mbox{\boldmath{$\theta$}}%
^{\mathbf{a}}})\nonumber\\
&  =\int%
\mbox{\boldmath{$\delta$}}%
\mbox{\boldmath{$\theta$}}%
^{\mathbf{a}}\wedge(-\star\mathcal{G}_{\mathbf{a}}+\frac{\partial
\mathfrak{L}_{m}}{\partial%
\mbox{\boldmath{$\theta$}}%
^{\mathbf{a}}})=0. \label{v15bis}%
\end{align}
and the equations of motion are:%
\begin{equation}
\mathcal{R}^{\mathbf{a}}-\frac{1}{2}R%
\mbox{\boldmath{$\theta$}}%
^{\mathbf{a}}=-\mathbf{T}^{\mathbf{a}}. \label{v16}%
\end{equation}

In Eq.(\ref{v16}) the $\mathcal{G}^{\mathbf{a}}=\left(  \mathcal{R}%
^{\mathbf{a}}-\frac{1}{2}R%
\mbox{\boldmath{$\theta$}}%
^{\mathbf{a}}\right)  \in\sec\bigwedge\nolimits^{1}T^{\ast}M$ $\hookrightarrow
\mathcal{C\ell}\left(  T^{\ast}M,\mathtt{g}\right)  $, $\mathcal{R}%
^{\mathbf{a}}=R_{\mathbf{b}}^{\mathbf{a}}%
\mbox{\boldmath{$\theta$}}%
^{\mathbf{b}}$ $\in\sec\bigwedge\nolimits^{1}T^{\ast}M\hookrightarrow
\mathcal{C\ell}\left(  T^{\ast}M,\mathtt{g}\right)  $, $R$ and the
$\mathbf{T}^{\mathbf{a}}$ have already been gave names. We moreover have:
\begin{equation}
\star\mathbf{T}^{\mathbf{a}}=-\star%
\slT
^{\mathbf{a}}:=-\frac{\partial\mathfrak{L}_{m}}{\partial%
\mbox{\boldmath{$\theta$}}%
^{\mathbf{a}}}\in\sec\bigwedge\nolimits^{3}T^{\ast}M, \label{v17}%
\end{equation}
as the definition of the energy-momentum $3$-forms of the matter fields.

We now have an important result, need for one of the purposes of the present
paper.\medskip

\textbf{Theorem }The\textbf{ }$\star\mathcal{G}^{\mathbf{a}}\in\sec
\bigwedge\nolimits^{3}T^{\ast}M$ $\hookrightarrow\mathcal{C\ell}\left(
T^{\ast}M,\mathtt{g}\right)  $ can be written:
\begin{equation}
-\star\mathcal{G}^{\mathbf{a}}=d\star\mathcal{S}^{\mathbf{a}}+\star
t^{\mathbf{a}}, \label{v18}%
\end{equation}
with
\begin{align}
\star\mathcal{S}^{\mathbf{c}}  &  =\frac{1}{2}\omega_{\mathbf{ab}}\wedge\star(%
\mbox{\boldmath{$\theta$}}%
^{\mathbf{a}}\wedge%
\mbox{\boldmath{$\theta$}}%
^{\mathbf{b}}\wedge%
\mbox{\boldmath{$\theta$}}%
^{\mathbf{c}})\in\sec\bigwedge\nolimits^{2}T^{\ast}M\label{19a}\\
\star t_{\mathbf{\ }}^{\mathbf{c}}  &  =-\frac{1}{2}\omega_{\mathbf{ab}}%
\wedge\lbrack\omega_{\mathbf{d}}^{\mathbf{c}}\wedge\star(%
\mbox{\boldmath{$\theta$}}%
^{\mathbf{a}}\wedge%
\mbox{\boldmath{$\theta$}}%
^{\mathbf{b}}\wedge%
\mbox{\boldmath{$\theta$}}%
^{\mathbf{d}})+\omega_{\mathbf{d}}^{\mathbf{b}}\wedge\star(%
\mbox{\boldmath{$\theta$}}%
^{\mathbf{a}}\wedge%
\mbox{\boldmath{$\theta$}}%
^{\mathbf{d}}\wedge%
\mbox{\boldmath{$\theta$}}%
^{\mathbf{c}})]\label{19}\\
&  \in\sec\bigwedge\nolimits^{3}T^{\ast}M.\nonumber
\end{align}
The $\star\mathcal{S}^{\mathbf{c}}\in\sec\bigwedge\nolimits^{2}T^{\ast}M$ are
called \textit{superpotentials} and the $\star t_{\mathbf{\ }}^{\mathbf{c}}$
are called the \textit{gravitational energy-momentum pseudo }$3$\textit{-
forms}. The reason for this name is given in Remark \textbf{1}.

\textbf{Proof}: To proof the theorem we compute $-2\star\mathcal{G}%
^{\mathbf{a}}$ as follows:%

\begin{align}
-  &  2\star\mathcal{G}^{\mathbf{d}}=d%
\mbox{\boldmath{$\omega$}}%
_{\mathbf{ab}}\wedge\star(%
\mbox{\boldmath{$\theta$}}%
^{\mathbf{a}}\wedge%
\mbox{\boldmath{$\theta$}}%
^{\mathbf{b}}\wedge%
\mbox{\boldmath{$\theta$}}%
^{\mathbf{d}})+%
\mbox{\boldmath{$\omega$}}%
_{\mathbf{ac}}\wedge%
\mbox{\boldmath{$\omega$}}%
_{\mathbf{b}}^{\mathbf{c}}\wedge\star(%
\mbox{\boldmath{$\theta$}}%
^{\mathbf{a}}\wedge%
\mbox{\boldmath{$\theta$}}%
^{\mathbf{b}}\wedge%
\mbox{\boldmath{$\theta$}}%
^{\mathbf{d}})\nonumber\\
&  =d[%
\mbox{\boldmath{$\omega$}}%
_{\mathbf{ab}}\wedge\star(%
\mbox{\boldmath{$\theta$}}%
^{\mathbf{a}}\wedge%
\mbox{\boldmath{$\theta$}}%
^{\mathbf{b}}\wedge%
\mbox{\boldmath{$\theta$}}%
^{\mathbf{d}})]+%
\mbox{\boldmath{$\omega$}}%
_{\mathbf{ab}}\wedge d\star(%
\mbox{\boldmath{$\theta$}}%
^{\mathbf{a}}\wedge%
\mbox{\boldmath{$\theta$}}%
^{\mathbf{b}}\wedge%
\mbox{\boldmath{$\theta$}}%
^{\mathbf{d}})\nonumber\\
&  +%
\mbox{\boldmath{$\omega$}}%
_{\mathbf{ac}}\wedge%
\mbox{\boldmath{$\omega$}}%
_{\mathbf{b}}^{\mathbf{c}}\wedge\star(%
\mbox{\boldmath{$\theta$}}%
^{\mathbf{a}}\wedge%
\mbox{\boldmath{$\theta$}}%
^{\mathbf{b}}\wedge%
\mbox{\boldmath{$\theta$}}%
^{\mathbf{d}})\nonumber\\
&  =d[%
\mbox{\boldmath{$\omega$}}%
_{\mathbf{ab}}\wedge\star(%
\mbox{\boldmath{$\theta$}}%
^{\mathbf{a}}\wedge%
\mbox{\boldmath{$\theta$}}%
^{\mathbf{b}}\wedge%
\mbox{\boldmath{$\theta$}}%
^{\mathbf{d}})]-%
\mbox{\boldmath{$\omega$}}%
_{\mathbf{ab}}\wedge%
\mbox{\boldmath{$\omega$}}%
_{\mathbf{p}}^{\mathbf{a}}\wedge\star(%
\mbox{\boldmath{$\theta$}}%
^{\mathbf{p}}\wedge%
\mbox{\boldmath{$\theta$}}%
^{\mathbf{b}}\wedge%
\mbox{\boldmath{$\theta$}}%
^{\mathbf{d}})\nonumber\\
&  -%
\mbox{\boldmath{$\omega$}}%
_{\mathbf{ab}}\wedge%
\mbox{\boldmath{$\omega$}}%
_{\mathbf{p}}^{\mathbf{b}}\wedge\star(%
\mbox{\boldmath{$\theta$}}%
^{\mathbf{a}}\wedge%
\mbox{\boldmath{$\theta$}}%
^{\mathbf{p}}\wedge%
\mbox{\boldmath{$\theta$}}%
^{\mathbf{d}})-%
\mbox{\boldmath{$\omega$}}%
_{\mathbf{ab}}\wedge%
\mbox{\boldmath{$\omega$}}%
_{\mathbf{p}}^{\mathbf{d}}\wedge\star(%
\mbox{\boldmath{$\theta$}}%
^{\mathbf{a}}\wedge%
\mbox{\boldmath{$\theta$}}%
^{\mathbf{b}}\wedge%
\mbox{\boldmath{$\theta$}}%
^{\mathbf{p}})\nonumber\\
&  +%
\mbox{\boldmath{$\omega$}}%
_{\mathbf{ac}}\wedge%
\mbox{\boldmath{$\omega$}}%
_{\mathbf{b}}^{\mathbf{c}}\wedge\star(%
\mbox{\boldmath{$\theta$}}%
^{\mathbf{a}}\wedge%
\mbox{\boldmath{$\theta$}}%
^{\mathbf{b}}\wedge%
\mbox{\boldmath{$\theta$}}%
^{\mathbf{d}})\nonumber\\
&  =d[%
\mbox{\boldmath{$\omega$}}%
_{\mathbf{ab}}\wedge\star(%
\mbox{\boldmath{$\theta$}}%
^{\mathbf{a}}\wedge%
\mbox{\boldmath{$\theta$}}%
^{\mathbf{b}}\wedge%
\mbox{\boldmath{$\theta$}}%
^{\mathbf{d}})]-%
\mbox{\boldmath{$\omega$}}%
_{\mathbf{ab}}\wedge\lbrack%
\mbox{\boldmath{$\omega$}}%
_{\mathbf{p}}^{\mathbf{d}}\wedge\star(%
\mbox{\boldmath{$\theta$}}%
^{\mathbf{a}}\wedge%
\mbox{\boldmath{$\theta$}}%
^{\mathbf{b}}\wedge%
\mbox{\boldmath{$\theta$}}%
^{\mathbf{p}})\nonumber\\
&  +%
\mbox{\boldmath{$\omega$}}%
_{\mathbf{p}}^{\mathbf{b}}\wedge\star(%
\mbox{\boldmath{$\theta$}}%
^{\mathbf{a}}\wedge%
\mbox{\boldmath{$\theta$}}%
^{\mathbf{p}}\wedge%
\mbox{\boldmath{$\theta$}}%
^{\mathbf{d}})]\nonumber\\
&  =2(d\star\mathcal{S}^{\mathbf{d}}+\star t^{\mathbf{d}}). \label{7.10.20}%
\end{align}

So, we just showed that Einstein equations can be written in the suggestive
form:%
\begin{equation}
-d\star\mathcal{S}^{\mathbf{a}}=(\star%
\slT
^{\mathbf{a}}+\star t^{\mathbf{a}}), \label{20}%
\end{equation}
which implies the differential conservation law $d(\star%
\slT
^{\mathbf{a}}+\star t^{\mathbf{a}})=0$, to be scrutinized below. We start,
with the\medskip

\noindent\textbf{Remark 1 }The $\star t_{\mathbf{\ }}^{\mathbf{a}}$ are not
true \textit{index} $3$-forms \cite{rodoliv2007}, i.e., there do not exist a
tensor field \ $\mathbf{t\in}\sec T_{1}^{3}M$ such that for $v_{i}\in\sec TM$,
$i=1,2,3,$
\begin{equation}
\mathbf{t}(v_{1},v_{2},v_{3},%
\mbox{\boldmath{$\theta$}}%
^{\mathbf{a}})=\star t^{\mathbf{a}}(v_{1},v_{2},v_{3}). \label{20bis}%
\end{equation}
We can immediately understand why this is the case, if we recall the
dependence of the $\star t^{\mathbf{a}}$ on the connection $1$-forms and that
these objects are \textit{gauge dependent} and thus do not transform
homogeneously under a change of orthonormal frame. Equivalently, the set
$t_{\mathbf{cd}}$ ($t_{\mathbf{c}}=t_{\mathbf{cd}}%
\mbox{\boldmath{$\theta$}}%
^{\mathbf{d}}$) for $\mathbf{c,d}=0,1,2,3$ are \textit{not} the components of
a tensor field. So, these components are said to define a
\textit{pseudo-tensor}.\medskip\ The $\star\mathcal{S}^{\mathbf{a}}$ also are
not true index forms for the same reason as the $\star t^{\mathbf{a}}$, they
are gauge dependent. \medskip

\noindent\textbf{Remark 2 \ }Eq.(\ref{20}) is known in recent literature of GR
as Sparling equations \cite{szabados} because it appears (in an equivalent
form) in a preprint \cite{sparling} of 1982 by that author. However, it
already appeared early, e.g., in a 1978 paper by Thirring and Wallner
\cite{thiwal}. \medskip

\noindent\textbf{Remark 3 \ }We emphasize that if we had used a coordinate
basis we would get analogous equations, i.e.%
\begin{equation}
-d\star\mathcal{S}^{\rho}=(\star%
\slT
^{\rho}+\star t^{\rho}), \label{21}%
\end{equation}%
\begin{align}
\star\mathcal{S}^{\rho}  &  =\frac{1}{2}\Gamma_{\alpha\beta}\wedge\star(%
\mbox{\boldmath{$\gamma$}}%
^{\alpha}\wedge%
\mbox{\boldmath{$\gamma$}}%
^{\beta}\wedge%
\mbox{\boldmath{$\gamma$}}%
^{\rho})\in\sec\bigwedge\nolimits^{2}T^{\ast}M\nonumber\\
\star t_{\mathbf{\ }}^{\rho}  &  =-\frac{1}{2}\Gamma_{\alpha\beta}%
\wedge\lbrack\Gamma_{\sigma}^{\rho}\wedge\star(%
\mbox{\boldmath{$\gamma$}}%
^{\alpha}\wedge%
\mbox{\boldmath{$\gamma$}}%
^{\beta}\wedge%
\mbox{\boldmath{$\gamma$}}%
^{\sigma})+\Gamma_{\sigma}^{\beta}\wedge\star(%
\mbox{\boldmath{$\gamma$}}%
^{\alpha}\wedge%
\mbox{\boldmath{$\gamma$}}%
^{\rho}\wedge%
\mbox{\boldmath{$\gamma$}}%
^{\sigma})]\label{22}\\
&  \in\sec\bigwedge\nolimits^{3}T^{\ast}M,\nonumber
\end{align}
with the $1$-form of connections given by $\mathbf{\Gamma}_{\sigma}^{\rho
}:=\Gamma_{\alpha\sigma}^{\rho}%
\mbox{\boldmath{$\gamma$}}%
^{\alpha}$, $D_{\partial_{\sigma}}%
\mbox{\boldmath{$\gamma$}}%
^{\rho}=-\Gamma_{\alpha\sigma}^{\rho}%
\mbox{\boldmath{$\gamma$}}%
^{\alpha}$.

Note that Eq.(\ref{21}), e.g., shows that each one of the $2$-form fields
$\star\mathcal{S}^{\mu}$ (\textit{the superpotentials}) is only defined modulo
a closed \ $2$-form \ $\star N^{\mu}$, $d\star N^{\mu}=0$.\medskip\ 

\noindent\textbf{Remark 4 \ }The use of a pseudo-tensor to express the
conservation law of energy-momentum of matter plus the gravitational field
appeared in a 1916 paper by Einstein \cite{einstein}. His pseudo-tensor which
has been originally presented in a coordinate basis are identified (using the
works of \cite{logunov} and \cite{trautman}) in the Appendix D. We show that
Einstein's superpotentials are the Freud's \textquotedblleft$2$-forms
\textquotedblright\ $\star\mathbf{U}^{\lambda}$ (Eq.(\ref{49b})).\medskip

\noindent\textbf{Remark 5 \ }We now turn to $%
\mbox{\boldmath{$\delta$}}%
{\displaystyle\int}
(\mathfrak{L}_{g}+\mathfrak{L}_{m})=0$. We immediately get%

\begin{equation}%
{\displaystyle\int}
\mbox{\boldmath{$\delta$}}%
\mbox{\boldmath{$\theta$}}%
^{\mathbf{a}}\left[  \frac{\partial\mathcal{L}_{g}}{\partial%
\mbox{\boldmath{$\theta$}}%
_{a}}+d\left(  \frac{\partial\mathcal{L}_{g}}{\partial d%
\mbox{\boldmath{$\theta$}}%
_{\mathbf{a}}}\right)  +\frac{\partial\mathfrak{L}_{m}}{\partial%
\mbox{\boldmath{$\theta$}}%
_{\mathbf{a}}}\right]  . \label{v3}%
\end{equation}
The computation of $\frac{\partial\mathcal{L}_{g}}{\partial%
\mbox{\boldmath{$\theta$}}%
_{a}}$ and $d\left(  \frac{\partial\mathcal{L}_{g}}{\partial d%
\mbox{\boldmath{$\theta$}}%
_{\mathbf{a}}}\right)  $ is a very long one and will not be given in this
paper. However, of course, we get:
\begin{equation}
-\star\mathcal{G}^{\mathbf{a}}=\frac{\partial\mathcal{L}_{g}}{\partial%
\mbox{\boldmath{$\theta$}}%
_{a}}+d\left(  \frac{\partial\mathcal{L}_{g}}{\partial d%
\mbox{\boldmath{$\theta$}}%
_{\mathbf{a}}}\right)  =\star\underset{g}{t}^{\mathbf{a}}+d\star\underset
{g}{\mathcal{S}}^{\mathbf{a}}=-\star%
\slT
^{\mathbf{a}}, \label{18}%
\end{equation}
and a detailed calculation (see details in \cite{rodoliv2007}) gives:
\begin{align}
\star\underset{g}{\mathcal{S}}^{\mathbf{a}}  &  =\frac{\partial\mathcal{L}%
_{g}}{\partial d%
\mbox{\boldmath{$\theta$}}%
_{\mathbf{a}}}=\star\mathcal{S}^{\mathbf{a}},\nonumber\\
\star\underset{g}{t}^{\mathbf{a}}  &  =\frac{\partial\mathcal{L}_{g}}{\partial%
\mbox{\boldmath{$\theta$}}%
_{a}}.=\star t^{\mathbf{a}}. \label{v19}%
\end{align}
\ \medskip

\section{Freud's Identity}

To compute the components of the $\mathcal{S}_{\mu}=\frac{1}{2}\mathcal{S}%
_{\mu}^{\nu\rho}%
\mbox{\boldmath{$\gamma$}}%
_{\nu}\wedge%
\mbox{\boldmath{$\gamma$}}%
_{\rho}\in\sec\bigwedge\nolimits^{2}T^{\ast}M$ is a trick exercise on the
algebra of differential forms. For that reason we give the details in the
Appendix \ C, where using the techniques of Clifford bundle formalism we found
directly that
\begin{equation}
\mathcal{S}_{\mu}^{\lambda\rho}=\frac{1}{2}\det%
\begin{bmatrix}
\delta_{\mu}^{\lambda} & \delta_{\mu}^{\sigma} & \delta_{\mu}^{\iota}\\
g^{\lambda\kappa} & g^{\sigma\kappa} & g^{\iota\kappa}\\
\Gamma_{\kappa\iota}^{\lambda} & \Gamma_{\kappa\iota}^{\sigma} & \Gamma
_{\iota\kappa}^{\iota}%
\end{bmatrix}
, \label{F}%
\end{equation}
which we moreover show to be equivalent to\footnote{We observe that
Eq.(\ref{23}) has also been found, e.g., in \cite{thiwal,trautman}.}
\begin{equation}
\mathcal{S}_{\mu}^{\nu\rho}=\frac{1}{2\sqrt{-\mathbf{g}}}\mathfrak{g}%
_{\mu\sigma}\partial_{\beta}(\mathfrak{g}^{\nu\beta}\mathfrak{g}^{\sigma\rho
}-\mathfrak{g}^{\rho\beta}\mathfrak{g}^{\sigma\nu}), \label{23}%
\end{equation}
with the definition of $\mathfrak{g}_{\mu\sigma}$ and $\mathfrak{g}^{\nu\beta
}$ given in Eq.(\ref{frackg}) and\ $\mathbf{g}$ in Eq.(\ref{det}) (Appendix A.1.1).

From Eq.(\ref{F})\ we immediately see (from the last formula in Freud's paper
\cite{freud}) that \ the object that he called $\mathfrak{U}_{\mu}^{\nu\rho}$
must be identified with\
\begin{equation}
\mathfrak{U}_{\mu}^{\nu\rho}=\sqrt{-\mathbf{g}}\mathcal{S}_{\mu}^{\nu\rho},
\label{24}%
\end{equation}
and the one he called $\mathfrak{U}_{\mu}^{\nu}$ \ (Eq.(1) of \cite{freud})
is
\begin{equation}
\mathfrak{U}_{\mu}^{\nu}=\frac{\partial}{\partial x^{\rho}}\mathfrak{U}_{\mu
}^{\nu\rho}. \label{25}%
\end{equation}

The $\mathfrak{U}_{\mu}^{\nu\rho}$ are the \textit{superpotentials} appearing
in Freud's classical paper and, of course,%
\begin{equation}
\frac{\partial}{\partial x^{\nu}}\mathfrak{U}_{\mu}^{\nu}=0.
\end{equation}
With the above identifications we verified in the Appendix that the identity
derived above (see Eq.(\ref{18}))
\begin{equation}
\mathcal{G}^{\iota}=-t^{\iota}-\star^{-1}d\star\mathcal{S}^{\iota}
\label{25b1}%
\end{equation}
is equivalent to%

\begin{align}
2\mathfrak{U}_{\kappa}^{\iota}  &  =\delta_{\kappa}^{\iota}\{\sqrt
{-\mathbf{g}}[R+g^{\mu\nu}\left(  \Gamma_{\mu\sigma}^{\rho}\Gamma_{\rho\nu
}^{\sigma}-\Gamma_{\mu\nu}^{\rho}\Gamma_{\rho\sigma}^{\sigma}\right)
]\}-2\sqrt{-\mathbf{g}}R_{\kappa}^{\iota}\nonumber\\
&  +\left(  \Gamma_{\mu\nu}^{\iota}\partial_{\varkappa}(\sqrt{-\mathbf{g}%
}g^{\mu\nu})-\Gamma_{\mu\nu}^{\nu}\partial_{\varkappa}(\sqrt{-\mathbf{g}%
}g^{\mu\iota})\right)  , \label{26}%
\end{align}
which is Eq.(8) in Freud's paper (Freud's identity) \cite{freud}.\medskip

In several papers and books \cite{san1,san2,santilli1,santilli2,santilli22}
\ Santilli claims that Einstein's gravitation in vacuum ($G_{\nu}^{\mu}=0)$ is
incompatible with the Freud identity of Riemannian geometry.

To endorse his claim, first Santilli printed a version of Freud's identity,
i.e.,\ his Eq.(3.10) in \cite{santilli2} (or in Eq.(1.4.10) in
\cite{santilli1}) with a \textit{missing term}, as we now show. Indeed,
putting $\mathfrak{R}_{\kappa}^{\iota}=\sqrt{-\mathbf{g}}R_{\kappa}^{\iota
},\;\mathfrak{R}=\sqrt{-\mathbf{g}}R$ and recalling the definition of
$\mathfrak{L}$ in Eq.(\ref{lag}), we can rewrite Eq.(\ref{26}) as:%
\begin{align}
&  \mathfrak{R}_{\kappa}^{\iota}-\frac{1}{2}\delta_{\kappa}^{\iota
}\mathfrak{R}-\frac{1}{2}\delta_{\kappa}^{\iota}\mathfrak{L}\nonumber\\
&  =\frac{1}{2}(\Gamma_{\mu\nu}^{\iota}\partial_{\varkappa}\mathfrak{g}%
^{\mu\nu}-\Gamma_{\mu\nu}^{\nu}\partial_{\varkappa}\mathfrak{g}^{\mu\iota
})-\mathfrak{U}_{\kappa}^{\iota}, \label{os1}%
\end{align}
Now, we can easily verify the identity\footnote{See page 70 of \cite{pauli}.}:%
\begin{equation}
\frac{1}{2}(\Gamma_{\mu\nu}^{\iota}\partial_{\varkappa}\mathfrak{g}^{\mu\nu
}-\Gamma_{\mu\nu}^{\nu}\partial_{\varkappa}\mathfrak{g}^{\mu\iota})=-\frac
{1}{2}\frac{\partial\mathfrak{L}}{\partial(\partial_{\iota}g^{\mu\nu}%
)}\partial_{\varkappa}g^{\mu\nu}, \label{os2}%
\end{equation}
which permit us to write%
\begin{align}
&  \mathfrak{R}_{\kappa}^{\iota}-\frac{1}{2}\delta_{\kappa}^{\iota
}\mathfrak{R}-\frac{1}{2}\delta_{\kappa}^{\iota}\mathfrak{L}\nonumber\\
&  =-\frac{1}{2}\frac{\partial\mathfrak{L}}{\partial(\partial_{\iota}g^{\mu
\nu})}\partial_{\varkappa}g^{\mu\nu}-\mathfrak{U}_{\kappa}^{\iota}.
\label{os3}%
\end{align}

This equation can also be writing, (with $\mathfrak{L}=\sqrt{-\mathbf{g}%
}\Theta$):%
\begin{align}
&  R_{\kappa}^{\iota}-\frac{1}{2}\delta_{\kappa}^{\iota}R-\frac{1}{2}%
\delta_{\kappa}^{\iota}\Theta\nonumber\\
&  =-\frac{1}{2\sqrt{-\mathbf{g}}}\frac{\partial\mathfrak{L}}{\partial
(\partial_{\iota}g^{\mu\nu})}\partial_{\varkappa}g^{\mu\nu}+\frac{1}%
{\sqrt{-\mathbf{g}}}\frac{\partial}{\partial x^{\rho}}\mathfrak{(}%
\sqrt{-\mathbf{g}}\mathfrak{\mathcal{S}_{\varkappa}^{\iota\rho})}, \label{os4}%
\end{align}
and since $\sqrt{-\mathbf{g}}$ does not depend on the $\partial_{\varkappa
}g^{\mu\nu}$ and $\partial_{\rho}\sqrt{-\mathbf{g}}=\Gamma_{\rho\sigma
}^{\sigma}\sqrt{-\mathbf{g}}$ we can still write:%
\begin{align}
&  R_{\kappa}^{\iota}-\frac{1}{2}\delta_{\kappa}^{\iota}R-\frac{1}{2}%
\delta_{\kappa}^{\iota}\Theta\nonumber\\
&  =-\frac{1}{2}\frac{\partial\Theta}{\partial(\partial_{\iota}g^{\mu\nu}%
)}\partial_{\varkappa}g^{\mu\nu}+\frac{\partial}{\partial x^{\rho}%
}\mathfrak{\mathcal{S}_{\varkappa}^{\iota\rho}}\nonumber\\
&  +\mathfrak{\mathcal{S}_{\varkappa}^{\iota\rho}}\Gamma_{\rho\sigma}^{\sigma
}, \label{os5}%
\end{align}
where $\mathfrak{\mathcal{S}_{\mu}^{\nu\rho}}$ is given by Eq.(\ref{n6}).
Eq.(\ref{os5}) can now be compared with Eq.(3.10) of \cite{santilli2} (or with
Eq.(1.4.10) in \cite{santilli1}) and we see that the last term, namely
$\mathfrak{\mathcal{S}_{\varkappa}^{\iota\rho}}\Gamma_{\rho\sigma}^{\sigma}$
is missing there\footnote{However, the equation printed in \cite{san1} is
correct.}.\smallskip

But leaving aside this "misprint", we then read, e.g., in \cite{santilli1} that:

{\small \textquotedblleft Therefore, the Freud identity requires two first
order source tensors for the exterior gravitational problems in vacuum, as in
Eq.(3.6.)\ of Ref.[1]\footnote{{\small Ref. [1] is the reference \cite{san1}
in the present paper.}}. These terms are absent in Einstein's gravitation
(1.4.1.)\footnote{{\small Eq.(1.4.1) in \cite{santilli1} is Einstein's field
equation without source, i.e., $G_{\mu\nu}=0$.}} that consequently, violates
the Freud identity of Riemannian geometry.\textquotedblleft}

First we must comment, that contrary to Santilli's statement, the two terms on
the right member of Eq.(\ref{os4}) are \textit{not} tensor fields, for indeed,
from \ what has been said above\textbf{\ }and taking into account
Eq.(\ref{18}) we know that Freud's identity is simply the component version of
a decomposition of the Einstein $3$-form fields $\star\mathcal{G}^{\mu}$ in
\textit{two} parts (one of then an exact differential), which however are
\textit{not} indexed forms, and thus are gauge dependent objects. Second, it
is necessary to become clear once and for ever that when $\star%
\slT
^{\mu}=0$, we simply have \ $-\star\mathcal{G}^{\mu}=d\star\mathcal{S}^{\mu
}+\star t^{\mu}=0$. , which is equivalent (Eq.(\ref{os5}) to:%

\begin{equation}
\mathfrak{R}_{\kappa}^{\iota}-\frac{1}{2}\delta_{\kappa}^{\iota}%
\mathfrak{R}=-\mathfrak{U}_{\kappa}^{\iota}+\frac{1}{2}\left(  \delta_{\kappa
}^{\iota}\mathfrak{L}-\frac{\partial\mathfrak{L}}{\partial(\partial_{\iota
}g^{\mu\nu})}\partial_{\varkappa}g^{\mu\nu}\right)  =0.
\end{equation}
What can be inferred from this equation is simply that the Ricci tensor of the
"exterior" problem is null.\footnote{We are not going to discuss here if the
exterior problem with a zero source term is a physically valid problem. We are
convinced that it is not, but certainly Santilli's proposed solution for that
problem inferred from his use of Freud's identity is not the answer to that
important issue.} And that is \textit{all}, there is \textit{no} inconsistency
between Einstein gravity, the Einstein-Hilbert field equations and Freud's identity.

\medskip

\noindent\textbf{Remark 6 } The fact that some people became confused during
decades with Freud's identity and its real meaning
\cite{alley,santilli1,santilli2,yi1,yi2,yi3} \ may certainly be attributed to
the use of the classical tensor calculus which, sometimes hides obvious things
for a long time. The identity, contrary to the hopes of
\cite{alley,yi1,yi2,yi3} does give a solution for the energy-momentum problem
in GR, even with the explicit introduction of an energy-momentum tensor for
the gravitational field, while maintaining that spacetime is a Lorentzian
manifold. The root of the problem consists in the obvious fact that there is
not even sense in GR to talk about the total energy momentum of particles
following different worldlines. The reason is crystal clear: in any manifold
not equipped with a teleparallel connection (as it is the case of a general
Lorentzian manifold, with non zero curvature tensor), we cannot sum vector
fields at different spacetime points. The problem of finding an
energy-momentum conservation law for matter fields in GR can be solved only in
a few special cases, namely when there exists appropriate Killing vector
fields in the Lorentzian manifold representing the gravitational field where
the matters fields generate and live (see, details, e.g., in \cite{rqr}%
).\medskip

\noindent\textbf{Remark 7 }We would also like to call the reader`s attention
to the fact that in \cite{san1} the quantity appearing in Definition II.11.3,%
\begin{equation}
R_{\kappa}^{\iota}-\frac{1}{2}\delta_{\kappa}^{\iota}R-\frac{1}{2}%
\delta_{\kappa}^{\iota}\Theta, \label{equiv}%
\end{equation}
is called the \textquotedblleft completed Einstein tensor\textquotedblright,
and it is stated that its covariant derivative is null. This statement is
wrong since the object given by Eq.(\ref{equiv}) is not a tensor. Indeed
although the two first terms define the Einstein tensor the term $\frac{1}%
{2}\delta_{\kappa}^{\iota}\Theta$ is \textit{not} a a tensor field. We observe
that already in 1916 Einstein at page 171 of the English translation of
\cite{einstein1} explicitly said that $\Theta$ is an invariant \textit{only}
with respect to linear transformations of coordinates, i.e., it is not a
scalar function in the manifold. Moreover, in a paper published in 1917
Levi-Civita , explicitly stated that $\Theta$ is \textit{not} a scalar
invariant \cite{levi} (see also \cite{schrodinger})\footnote{By the way, a
proof that $\Theta$ is not a scalar is as follows. Calculate its value at a
given point spacetime point using arbitrary coordinates. You get in general
that $\Theta$ is non null (you can verify this with an example, e.g., using
the Schwarzschild in standard coordinates). Next introduce Riemann normal
coordinates covering that spacetime point. Using these coordinates all
connection are zero at that point and then the evaluation of $\Theta$ now
gives zero.}. And, since $\frac{1}{2}\delta_{\kappa}^{\iota}\Theta$ is not a
tensor field there is no meaning in taking its covariant derivative, and
consequently Corollary II.11.2.1 in \cite{san1} is false.

\section{Freud's Identity and the Energy-Momentum \textquotedblleft
Conservation Law\textquotedblright\ of GR}

We already comment that Freud's identity through Eq.(\ref{20}) (or
Eq.(\ref{21})) suggests that we have found a conservation law for the
energy-momentum of matter plus the gravitational field in GR. Indeed, from
Eq.(\ref{21}), it follows that \
\begin{equation}
d(\star%
\slT
^{\mu}+\star t^{\mu})=0. \label{29}%
\end{equation}

However, this is simply a wishful thinking, since \ the $\star t^{\mu}$ are
gauge dependent quantities and that fact implies that one of the definitions
of the "inertial' mass of the source, in GR given by \cite{thiwal}
\begin{equation}
m_{\mathbf{I}}=-%
{\displaystyle\int\nolimits_{V}}
\star(%
\slT
^{0}+t^{0})=%
{\displaystyle\int\nolimits_{\partial V}}
\star\mathcal{S}^{0} \label{30}%
\end{equation}
takes a value that depends on the coordinate system that we choose to make the computation.

In truth, Eq.(\ref{30}), printed in many papers and books results from a naive
use of Stokes theorem. Indeed, such a theorem is valid one for the integration
of \textit{true} differential forms (under well known conditions). If we
recall the well known definition of the integral of a differential form
\cite{choquet,felsager} we see that a coordinate free result depends
fundamentally on the fact that the differential form being integrated defines
a \textit{true} tensor. However, as already mentioned in Remark \textbf{1},
the $\star\mathcal{S}^{\mu}$ are not true indexed forms, and so their
integration will be certainly coordinate dependent \cite{boro}. In Appendix C
for completeness and hopping that the present paper may be of some utility for
people trying to understand this issue, we find also from our formalism the so
called Einstein and the Landau-Lifshitz \textquotedblleft
inertial\textquotedblright\ masses (concepts which have the same problems as
the one defined in Eq.(\ref{30})).\medskip

The problem just discussed is a really serious one if we take GR as a valid
theory of the gravitational field, for it means that in that theory there are
no conservation laws of energy-momentum (and also of angular momentum) despite
almost 100 years of hard work by several people \footnote{A detailed
discussion of conservation laws in a general Riemann-Cartan spacetime is given
\cite{rqr}.}. And, at this point it is better to quote page 98 of Sachs\&Wu
\cite{sawu}:

{\footnotesize \ }{\small \textquotedblleft\ As mentioned in section 3.8,
conservation laws have a great predictive power. It is a shame to lose the
special relativistic total energy conservation law (Section 3.10.2) in general
relativity. Many of the attempts to resurrect it are quite interesting; many
are simply garbage.\textquotedblright}

\section{Conclusions}

In this paper we proved that contrary to the claim in
\cite{santilli1,santilli2}, there is no incompatibility from the mathematical
point of view between Freud's identity and Einstein-Hilbert field equations of
GR, both in vacuum and inside matter. Freud's identity, or disguised versions
of it, have been used by several people during all XX$^{th}$ century to try to
give a meaning to conservation laws of energy-momentum and angular momentum in
GR. These efforts unfortunately resulted in no success, of course, because
Freud's identity involves the use of pseudo-tensors (something that is
absolutely obvious in our presentation), and thus gives global quantities
(i.e., the result of integrals) depending of the coordinate chart used (see
also Appendices D and E). This is a serious and vexatious problem that we
believe, will need a radical change of paradigm to be solved\footnote{Using
the asymptotic flatness notions, first introduced by Penrose \cite{penrose},
it is possible for some "isolated systems" to introduce the ADM and the Bondi
masses. It is even possibel to prove that the Bondi mass is positive
\cite{witten} \ But even if the notion of Bondi mass is considered by many a
good solution to the nergy-problem in GR, the fact is that it did not solve
the problem in principle. It is only a calculational device. n introduction to
asymptopita, ADM and Bondi masses can be found in \cite{stewart}.\ }. As
discussed in, e.g., \cite{notte,rodoliv2007} a possible solution (maintaining
the Einstein-Hilbert equations in an appropriate form) can be given with the
gravitational field interpreted as field in Faraday sense living in Minkowski
spacetime (or other background spacetime equipped with absolute
parallelism)\footnote{Recently Gorelik proposed in an interesting paper
\cite{gorelik} to use the \textit{quasi Poincar\'{e} group} of a Riemannian
space as the generator of the Noether symmetries leading to conservation laws
of "energy-momentum", angular momentum and "center of mass motion". A need
comment on this approach that do not involve the use of the Freud's identity
will be presented somewhere.}. The geometrical interpretation of gravitation
as " geometry of spacetime" is a simple coincidence \cite{notte,weinberg}
(valid only to a certain degree of approximation).

\appendix

\section{Clifford Bundle Formalism}

Let $\mathfrak{M}=(M,%
\slg
,D,\tau_{g},\uparrow)$ be an arbitrary Lorentzian spacetime.\ The quadruple
$(M,%
\slg
,\tau_{g},\uparrow)$ denotes a four-dimensional time-oriented and
space-oriented Lorentzian manifold \cite{rodoliv2007,sawu}. This means that $%
\slg
\in\sec T_{2}^{0}M$ is a Lorentzian metric of signature (1,3), $\tau_{g}%
\in\sec\bigwedge{}^{4}T^{\ast}M$ and $\uparrow$ is a time-orientation (see
details, e.g., in \cite{sawu}). Here, $T^{\ast}M$ [$TM$] is the cotangent
[tangent] bundle. $T^{\ast}M=\cup_{x\in M}T_{x}^{\ast}M$, $TM=\cup_{x\in
M}T_{x}M$, and $T_{x}M\simeq T_{x}^{\ast}M\simeq\mathbb{R}^{1,3}$, where
$\mathbb{R}^{1,3}$ is the Minkowski vector space\footnote{Not to be confused
with Minkowski spacetime \cite{sawu}.}. $D$ is the Levi-Civita connection of $%
\slg
$, i.e., it is a\ metric compatible connection, which implies $D%
\slg
=0$. In general, $\mathbf{R}=\mathbf{R}^{D}\neq0$, $\Theta=\Theta^{D}=0$,
$\mathbf{R}$ and $\Theta$ being respectively the curvature and torsion tensors
of the connection. Minkowski spacetime is the particular case of a Lorentzian
spacetime for which $\mathbf{R}=0$, $\Theta=0$, and $M\simeq\mathbb{R}^{4}$.
Let $\mathtt{g}\in\sec T_{0}^{2}M$ be the metric of the \textit{cotangent
bundle}. The Clifford bundle of differential forms $\mathcal{C}\!\ell
(M,\mathtt{g})$ is the bundle of algebras, i.e., $\mathcal{C}\ell
(M,\mathtt{g})=\cup_{x\in M}\mathcal{C}\!\ell(T_{x}^{\ast}M,\mathtt{g})$,
where $\forall x\in M$, $\mathcal{C}\!\ell(T_{x}^{\ast}M,\mathtt{g}%
)=\mathbb{R}_{1,3}$, the so called \emph{spacetime} \emph{algebra
}\cite{rodoliv2007}. Recall also that $\mathcal{C}\!\ell(M,\mathtt{g})$ is a
vector bundle associated to the \emph{orthonormal frame bundle}, i.e.,
$\mathcal{C}\!\ell(M,\mathtt{g})$ $=P_{\mathrm{SO}_{(1,3)}^{e}}(M)\times
_{\mathrm{Ad}}\mathcal{C}l_{1,3}$ \cite{lawmi,moro}. For any $x\in M$,
$\mathcal{C}\!\ell(T_{x}^{\ast}M,\left.  \mathtt{g}\right\vert _{x})$ as a
linear space over the real field $\mathbb{R}$ is isomorphic to the Cartan
algebra $\bigwedge T_{x}^{\ast}M$ of the cotangent space. $\bigwedge
T_{x}^{\ast}M=\oplus_{k=0}^{4}\bigwedge^{k}T_{x}^{\ast}M$, where
$\bigwedge^{k}T_{x}^{\ast}M$ is the $\binom{4}{k}$-dimensional space of
$k$-forms. Then, sections of $\mathcal{C}\!\ell(M,\mathtt{g})$ can be
represented as a sum of non homogeneous differential forms, that will be
called Clifford (multiform) fields. In the Clifford bundle formalism, of
course, arbitrary basis can be used (see remark below), but in this short
review of the main ideas of the Clifford calculus we use orthonormal basis.
Let then $\{\mathbf{e}_{\mathbf{a}}\}$ be an orthonormal basis for $TU\subset
TM$, i.e., $\mathtt{g}(\mathbf{e}_{\mathbf{a}},\mathbf{e}_{\mathbf{a}}%
)=\eta_{\mathbf{ab}}=\mathrm{diag}(1,-1,-1,-1)$. Let $%
\mbox{\boldmath{$\theta$}}%
^{\mathbf{a}}\in\sec\bigwedge^{1}T^{\ast}M\hookrightarrow\sec\mathcal{C}%
\!\ell(M,\mathtt{g})$ ($\mathbf{a}=0,1,2,3$) be such that the set $\{%
\mbox{\boldmath{$\theta$}}%
^{\mathbf{a}}\}$ is the dual basis of $\{\mathbf{e}_{\mathbf{a}}\}$.

\subsection{Clifford Product}

The fundamental \emph{Clifford product} (in what follows to be denoted by
juxtaposition of symbols) is generated by
\begin{equation}%
\mbox{\boldmath{$\theta$}}%
^{\mathbf{a}}%
\mbox{\boldmath{$\theta$}}%
^{\mathbf{b}}+%
\mbox{\boldmath{$\theta$}}%
^{\mathbf{b}}%
\mbox{\boldmath{$\theta$}}%
^{\mathbf{a}}=2\eta^{\mathbf{ab}} \label{cl}%
\end{equation}
and if $\mathcal{C}\in\sec\mathcal{C}\ell(M,\mathtt{g})$ we have%

\begin{equation}
\mathcal{C}=s+v_{\mathbf{a}}^{\mathbf{a}}%
\mbox{\boldmath{$\theta$}}%
+\frac{1}{2!}f_{\mathbf{ab}}%
\mbox{\boldmath{$\theta$}}%
^{\mathbf{a}}%
\mbox{\boldmath{$\theta$}}%
^{\mathbf{b}}+\frac{1}{3!}t_{\mathbf{abc}}%
\mbox{\boldmath{$\theta$}}%
^{\mathbf{a}}%
\mbox{\boldmath{$\theta$}}%
^{\mathbf{b}}%
\mbox{\boldmath{$\theta$}}%
^{\mathbf{c}}+p%
\mbox{\boldmath{$\theta$}}%
^{5}\;, \label{3}%
\end{equation}
where $\tau_{g}=%
\mbox{\boldmath{$\theta$}}%
^{5}=%
\mbox{\boldmath{$\theta$}}%
^{0}%
\mbox{\boldmath{$\theta$}}%
^{\mathbf{1}}%
\mbox{\boldmath{$\theta$}}%
^{\mathbf{2}}%
\mbox{\boldmath{$\theta$}}%
^{\mathbf{3}}$ is the volume element and $s$, $v_{\mathbf{a}}$,
$f_{\mathbf{ab}}$, $t_{\mathbf{abc}}$, $p\in\sec\bigwedge^{0}T^{\ast
}M\hookrightarrow\sec\mathcal{C}\!\ell(M,\mathtt{g})$.

For $A_{r}\in\sec\bigwedge^{r}T^{\ast}M\hookrightarrow\sec\mathcal{C}%
\!\ell(M,\mathtt{g}),B_{s}\in\sec\bigwedge^{s}T^{\ast}M\hookrightarrow
\sec\mathcal{C}\!\ell(M,\mathtt{g})$ we define the \emph{exterior product} in
$\mathcal{C}\!\ell(M,\mathtt{g})$ \ ($\forall r,s=0,1,2,3)$ by
\begin{equation}
A_{r}\wedge B_{s}=\langle A_{r}B_{s}\rangle_{r+s}, \label{5}%
\end{equation}
where $\langle\;\;\rangle_{k}$ is the component in $\bigwedge^{k}T^{\ast}M$ of
the Clifford field. Of course, $A_{r}\wedge B_{s}=(-1)^{rs}B_{s}\wedge A_{r}$,
and the exterior product is extended by linearity to all sections of
$\mathcal{C}\!\ell(M,\mathtt{g})$.

Let $A_{r}\in\sec\bigwedge^{r}T^{\ast}M\hookrightarrow\sec\mathcal{C}%
\!\ell(M,\mathtt{g}),B_{s}\in\sec\bigwedge^{s}T^{\ast}M\hookrightarrow
\sec\mathcal{C}\!\ell(M,\mathtt{g})$. We define a \emph{scalar product
}in\emph{\ }$\mathcal{C}\!\ell(M,\mathtt{g})$ (denoted by $\cdot$) as follows:

(i) For $a,b\in\sec\bigwedge^{1}T^{\ast}M\hookrightarrow\sec\mathcal{C}%
\!\ell(M,\mathtt{g}),$%
\begin{equation}
a\cdot b=\frac{1}{2}(ab+ba)=\mathtt{g}(a,b). \label{4}%
\end{equation}

(ii) For $A_{r}=a_{1}\wedge...\wedge a_{r},B_{r}=b_{1}\wedge...\wedge b_{r}$,
$a_{i},b_{j}\in\sec\bigwedge^{1}T^{\ast}M\hookrightarrow\sec\mathcal{C}%
\!\ell(M,\mathtt{g})$, $i,j=1,...,r,$
\begin{align}
A_{r}\cdot B_{r}  &  =(a_{1}\wedge...\wedge a_{r})\cdot(b_{1}\wedge...\wedge
b_{r})\nonumber\\
&  =\left\vert
\begin{array}
[c]{lll}%
a_{1}\cdot b_{1} & .... & a_{1}\cdot b_{r}\\
.......... & .... & ..........\\
a_{r}\cdot b_{1} & .... & a_{r}\cdot b_{r}%
\end{array}
\right\vert . \label{6}%
\end{align}

We agree that if $r=s=0$, the scalar product is simply the ordinary product in
the real field.

Also, if $r\neq s$, then $A_{r}\cdot B_{s}=0$. Finally, the scalar product is
extended by linearity for all sections of $\mathcal{C}\!\ell(M,\mathtt{g})$.

For $r\leq s$, $A_{r}=a_{1}\wedge...\wedge a_{r}$, $B_{s}=b_{1}\wedge...\wedge
b_{s\text{ }}$, we define the \textit{left contraction} $\lrcorner
:(A_{r},B_{s})\mapsto A_{r}\lrcorner B_{s}$ by
\begin{equation}
A_{r}\lrcorner B_{s}=%
{\displaystyle\sum\limits_{i_{1}\,<...\,<i_{r}}}
\epsilon^{i_{1}...i_{s}}(a_{1}\wedge...\wedge a_{r})\cdot(b_{_{i_{1}}}%
\wedge...\wedge b_{i_{r}})^{\sim}b_{i_{r}+1}\wedge...\wedge b_{i_{s}}
\label{7}%
\end{equation}
where $\sim$ is the reverse mapping (\emph{reversion}) defined by
\begin{align}
\symbol{126}  &  :\sec\mathcal{C}\!\ell(M,\mathtt{g})\rightarrow
\sec\mathcal{C}\!\ell(M,\mathtt{g}),\nonumber\\
\tilde{A}  &  =%
{\displaystyle\sum\limits_{p=0}^{4}}
\text{ }\tilde{A}_{p}=%
{\displaystyle\sum\limits_{p=0}^{4}}
(-1)^{\frac{1}{2}k(k-1)}A_{p},\nonumber\\
A_{p}  &  \in\sec%
{\displaystyle\bigwedge\nolimits^{p}}
T^{\ast}M\hookrightarrow\sec\mathcal{C}\!\ell(M,\mathtt{g}).
\end{align}
We agree that for $\alpha,\beta\in\sec\bigwedge^{0}T^{\ast}M$ the contraction
is the ordinary (pointwise) product in the real field and that if $\alpha
\in\sec\bigwedge^{0}T^{\ast}M$, $A_{r}\in\sec\bigwedge^{r}T^{\ast}M,B_{s}%
\in\sec\bigwedge^{s}T^{\ast}M\hookrightarrow\sec\mathcal{C}\!\ell
(M,\mathtt{g})$ then $(\alpha A_{r})\lrcorner B_{s}=A_{r}\lrcorner(\alpha
B_{s})$. Left contraction is extended by linearity to all pairs of sections of
$\mathcal{C}\!\ell(M,\mathtt{g})$, i.e., for $A,B\in\sec\mathcal{C}%
\!\ell(M,\mathtt{g})$%
\begin{equation}
A\lrcorner B=\sum_{r,s}\langle A\rangle_{r}\lrcorner\langle B\rangle_{s},\quad
r\leq s. \label{9}%
\end{equation}

It is also necessary to introduce the operator of \emph{right contraction}
denoted by $\llcorner$. The definition is obtained from the one presenting the
left contraction with the imposition that $r\geq s$ and taking into account
that now if $A_{r}\in\sec\bigwedge^{r}T^{\ast}M,$ $B_{s}\in\sec\bigwedge
^{s}T^{\ast}M$ then $A_{r}\llcorner(\alpha B_{s})=(\alpha A_{r})\llcorner
B_{s}$. See also the third formula in Eq.(\ref{10}).

The main formulas used in this paper can be obtained from the following ones
\begin{align}
a\mathcal{B}_{s}  &  =a\lrcorner\mathcal{B}_{s}+a\wedge\mathcal{B}%
_{s},\;\;\mathcal{B}_{s}a=\mathcal{B}_{s}\llcorner a+\mathcal{B}_{s}\wedge
a,\nonumber\\
a\lrcorner\mathcal{B}_{s}  &  =\frac{1}{2}(a\mathcal{B}_{s}-(-1)^{s}%
\mathcal{B}_{s}a),\nonumber\\
\mathcal{A}_{r}\lrcorner\mathcal{B}_{s}  &  =(-1)^{r(s-r)}\mathcal{B}%
_{s}\llcorner\mathcal{A}_{r},\nonumber\\
a\wedge\mathcal{B}_{s}  &  =\frac{1}{2}(a\mathcal{B}_{s}+(-1)^{s}%
\mathcal{B}_{s}a),\nonumber\\
\mathcal{A}_{r}\mathcal{B}_{s}  &  =\langle\mathcal{A}_{r}\mathcal{B}%
_{s}\rangle_{|r-s|}+\langle\mathcal{A}_{r}\mathcal{B}_{s}\rangle
_{|r-s|+2}+...+\langle\mathcal{A}_{r}\mathcal{B}_{s}\rangle_{|r+s|}\nonumber\\
&  =\sum\limits_{k=0}^{m}\langle\mathcal{A}_{r}\mathcal{B}_{s}\rangle
_{|r-s|+2k}\text{ }\nonumber\\
\mathcal{A}_{r}\cdot\mathcal{B}_{r}  &  =\mathcal{B}_{r}\cdot\mathcal{A}%
_{r}=\widetilde{\mathcal{A}}_{r}\text{ }\lrcorner\mathcal{B}_{r}%
=\mathcal{A}_{r}\llcorner\widetilde{\mathcal{B}}_{r}=\langle\widetilde
{\mathcal{A}}_{r}\mathcal{B}_{r}\rangle_{0}=\langle\mathcal{A}_{r}%
\widetilde{\mathcal{B}}_{r}\rangle_{0}. \label{10}%
\end{align}
Two other important identities to be used below are:%

\begin{equation}
a\lrcorner(\mathcal{X}\wedge\mathcal{Y})=(a\lrcorner\mathcal{X})\wedge
\mathcal{Y}+\mathcal{\hat{X}}\wedge(a\lrcorner\mathcal{Y}), \label{T54}%
\end{equation}
for any $a\in\sec%
{\displaystyle\bigwedge\nolimits^{1}}
T^{\ast}M$ and $\mathcal{X},\mathcal{Y}\in\sec%
{\displaystyle\bigwedge}
T^{\ast}M$, and
\begin{equation}
A\lrcorner(B\lrcorner C)=(A\wedge B)\lrcorner C, \label{T50}%
\end{equation}
for any $A,B,C\in\sec\bigwedge T^{\ast}M\hookrightarrow\mathcal{C}%
\ell(M,\mathtt{g})$.

\subsubsection{Hodge Star Operator}

Let $\star$ be the Hodge star operator, i.e., the mapping
\begin{equation}
\star:%
{\displaystyle\bigwedge\nolimits^{k}}
T^{\ast}M\rightarrow%
{\displaystyle\bigwedge\nolimits^{4-k}}
T^{\ast}M,\text{ }A_{k}\mapsto\star A_{k} \label{h}%
\end{equation}
where for $A_{k}\in\sec\bigwedge^{k}T^{\ast}M\hookrightarrow\sec
\mathcal{C}\!\ell(M,\mathtt{g})$%
\begin{equation}
\lbrack B_{k}\cdot A_{k}]\tau_{g}=B_{k}\wedge\star A_{k},\forall B_{k}\in
\sec\bigwedge\nolimits^{k}T^{\ast}M\hookrightarrow\sec\mathcal{C}%
\!\ell(M,\mathtt{g}). \label{11a}%
\end{equation}
$\tau_{\mathtt{g}}=\theta^{\mathbf{5}}\in\sec\bigwedge^{4}T^{\ast
}M\hookrightarrow\sec\mathcal{C}\!\ell(M,\mathtt{g})$ is a \emph{standard}
volume element. Then we can easily verify that
\begin{equation}
\star A_{k}=\widetilde{A}_{k}\tau_{\mathtt{g}}=\widetilde{A}_{k}\lrcorner
\tau_{\mathtt{g}}. \label{11b}%
\end{equation}
where as noted before, in this paper $\widetilde{\mathcal{A}}_{k}$ denotes the
\textit{reverse} of $\mathcal{A}_{k}$. Eq.(\ref{11b}) permits calculation of
Hodge duals very easily in an orthonormal basis for which $\tau_{\mathtt{g}}=%
\mbox{\boldmath{$\theta$}}%
^{\mathbf{5}}$. Let $\{\vartheta^{\alpha}\}$ be the dual basis of
$\{e_{\alpha}\}$ (i.e., it is a basis for $T^{\ast}U\equiv\bigwedge
\nolimits^{1}T^{\ast}U$) which is either orthonormal or a coordinate basis.
Then writing \texttt{g}$(\vartheta^{\alpha},\vartheta^{\beta})=g^{\alpha\beta
}$, with $g^{\alpha\beta}g_{\alpha\rho}=\delta_{\rho}^{\beta}$, and
$\vartheta^{\mu_{1}...\mu_{p}}=\vartheta^{\mu_{1}}\wedge...\wedge
\vartheta^{\mu_{p}}$, $\vartheta^{\nu_{p+1}...\nu_{n}}=\vartheta^{\nu_{p+1}%
}\wedge...\wedge\vartheta^{\nu_{n}}$ we have from Eq.(\ref{11b})
\begin{equation}
{}\star\theta^{\mu_{1}...\mu_{p}}=\frac{1}{(n-p)!}\sqrt{\left\vert
\mathbf{g}\right\vert }g^{\mu_{1}\nu_{1}}...g^{\mu_{p}\nu_{p}}\epsilon
_{\nu_{1}...\nu_{n}}\vartheta^{\nu_{p+1}...\nu_{n}}. \label{hodge dual}%
\end{equation}
where $\mathbf{g}$ denotes the determinant of the matrix with entries
$g_{\alpha\beta}=$\texttt{ }$%
\slg
(e_{\alpha},e_{\beta})$, i.e.,
\begin{equation}
\mathbf{g}=\det[g_{\alpha\beta}]. \label{det}%
\end{equation}
We also define the inverse $\star^{-1}$ of the Hodge dual operator, such that
\ $\star^{-1}\star=\star\star^{-1}=1$. It is given by:
\begin{align}
\star^{-1}  &  :\sec%
{\displaystyle\bigwedge\nolimits^{n-r}}
T^{\ast}M\rightarrow\sec%
{\displaystyle\bigwedge\nolimits^{r}}
T^{\ast}M,\nonumber\\
\star^{-1}  &  =(-1)^{r(n-r)}\mathrm{sgn}\text{ }\mathbf{g\,}\star, \label{h1}%
\end{align}
where \textrm{sgn }$\mathbf{g}=\mathbf{g}/|\mathbf{g}|$ denotes the sign of
the determinant $\mathbf{g}$.

Some useful identities (used several times below) involving the Hodge star
operator, the exterior product and contractions are:%

\begin{equation}%
\begin{array}
[c]{l}%
A_{r}\wedge\star B_{s}=B_{s}\wedge\star A_{r};\quad r=s\\
A_{r}\cdot\star B_{s}=B_{s}\cdot\star A_{r};\quad r+s=n\\
A_{r}\wedge\star B_{s}=(-1)^{r(s-1)}\star(\tilde{A}_{r}\lrcorner B_{s});\quad
r\leq s\\
A_{r}\lrcorner\star B_{s}=(-1)^{rs}\star(\tilde{A}_{r}\wedge B_{s});\quad
r+s\leq n\\
\star\tau_{g}=\mathrm{sign}\text{ }\mathbf{g};\quad\star1=\tau_{g}.
\end{array}
\label{440new}%
\end{equation}

\subsubsection{Dirac Operator Associated to a Levi-Civita Connection}

Let $d$ and $\delta$ be respectively the differential and Hodge codifferential
operators acting on sections of $\mathcal{C}\!\ell(M,\mathtt{g})$. If
$A_{p}\in\sec\bigwedge^{p}T^{\ast}M\hookrightarrow\sec\mathcal{C}%
\!\ell(M,\mathtt{g})$, then $\delta A_{p}=(-1)^{p}\star^{-1}d\star A_{p}$.

The Dirac operator acting on sections of $\mathcal{C}\!\ell(M,\mathtt{g})$
associated with the metric compatible connection $D$ is the invariant first
order differential operator
\begin{equation}
{\mbox{\boldmath$\partial$}}=%
\mbox{\boldmath{$\theta$}}%
^{\mathbf{a}}D_{\mathbf{e}_{\mathbf{a}}}, \label{12}%
\end{equation}
where $\{\mathbf{e}_{\mathbf{a}}\}$ is an arbitrary \emph{orthonormal basis}
for $TU\subset TM$ and $\{%
\mbox{\boldmath{$\theta$}}%
^{\mathbf{b}}\}$ is a basis for $T^{\ast}U\subset T^{\ast}M$ dual to the basis
$\{\mathbf{e}_{\mathbf{a}}\}$, i.e., $%
\mbox{\boldmath{$\theta$}}%
^{\mathbf{b}}(\mathbf{e}_{\mathbf{a}})=\delta_{\mathbf{b}}^{\mathbf{a}}$,
$\mathbf{a,b}=0,1,2,3$. The reciprocal basis of $\{%
\mbox{\boldmath{$\theta$}}%
^{\mathbf{b}}\}$ is denoted $\{\theta_{\mathbf{a}}\}$ and we have
$\theta_{\mathbf{a}}\cdot\theta_{\mathbf{b}}=\eta_{\mathbf{ab}}$. Also,
\begin{equation}
D_{\mathbf{e}_{\mathbf{a}}}%
\mbox{\boldmath{$\theta$}}%
^{\mathbf{b}}=-\omega_{\mathbf{a}}^{\mathbf{bc}}%
\mbox{\boldmath{$\theta$}}%
_{\mathbf{c}} \label{12n}%
\end{equation}
and we write the connection $1$-forms in the orthogonal gauge as%
\begin{equation}
\omega_{\mathbf{b}}^{\mathbf{a}}:=\omega_{\mathbf{cb}}^{\mathbf{a}}%
\mbox{\boldmath{$\theta$}}%
^{\mathbf{c}}. \label{12na}%
\end{equation}
Moreover, we introduce the objects $\mathbf{\omega}_{\mathbf{e}_{\mathbf{a}}%
}\in\sec\bigwedge^{2}T^{\ast}M,$
\begin{equation}
\mathbf{\omega}_{\mathbf{e}_{\mathbf{a}}}=\frac{1}{2}\omega_{\mathbf{a}%
}^{\mathbf{bc}}%
\mbox{\boldmath{$\theta$}}%
_{\mathbf{b}}\wedge%
\mbox{\boldmath{$\theta$}}%
_{\mathbf{c}}. \label{12nn}%
\end{equation}
Then, for any $A_{p}\in\sec\bigwedge^{p}T^{\ast}M,$ $p=0,1,2,3,4$ we can
write
\begin{equation}
D_{\mathbf{e}_{\mathbf{a}}}A_{p}=\partial_{\mathbf{e}_{\mathbf{a}}}A_{p}%
+\frac{1}{2}[\mathbf{\omega}_{\mathbf{e}_{\mathbf{a}}},A_{p}], \label{12nnn}%
\end{equation}
where $\partial_{\mathbf{e}_{\mathbf{a}}}$ is the Pfaff derivative, i.e., if
$A_{p}=\frac{1}{p!}A_{\mathbf{i}_{1}...\mathbf{i}_{p}}\theta^{\mathbf{i}%
_{1}...\mathbf{i}_{p}}$,%
\begin{equation}
\partial_{\mathbf{e}_{\mathbf{a}}}A_{p}:=\frac{1}{p!}\mathbf{e}_{\mathbf{a}%
}(A_{\mathbf{i}_{1}...\mathbf{i}_{p}})%
\mbox{\boldmath{$\theta$}}%
^{\mathbf{i}_{1}...\mathbf{i}_{p}}. \label{pfaaf}%
\end{equation}

Eq.(\ref{12nnn}) is an important formula which is also valid for a
nonhomogeneous $A\in\sec\mathcal{C}\ell(M,\mathtt{g})$. It is proved, e.g., in
\cite{moro,rodoliv2007}.

We have also the important result:%

\begin{align}
{\mbox{\boldmath$\partial$}}A_{p}  &  ={\mbox{\boldmath$\partial$}}\wedge
A_{p\,}+\,{\mbox{\boldmath$\partial$}}\lrcorner A_{p}=dA_{p}-\delta
A_{p},\nonumber\\
{\mbox{\boldmath$\partial$}}\wedge A_{p}  &  =dA_{p},\hspace{0.1in}%
\,{\mbox{\boldmath$\partial$}}\lrcorner A_{p}=-\delta A_{p}. \label{13}%
\end{align}

\noindent\textbf{Remark 8. }We conclude this section by emphasizing that the
formalism just presented is valid in an arbitrary coordinate basis
\ $\{{\mbox{\boldmath$\partial$}}_{\mu}\}$ of $TU\subset TM$ associated to
local coordinates $\{x^{\mu}\}$ covering $U$. In this case if \ $\{%
\mbox{\boldmath{$\theta$}}%
^{\mu}=dx^{\mu}\}$ is the dual basis of $\{{\mbox{\boldmath$\partial$}}_{\mu
}\}$ we write
\begin{equation}
D_{\partial_{\mu}}\partial_{\nu}=\Gamma_{\mu\nu}^{\rho}\partial_{\rho}\text{
\ \ \ \ \ \ \ \ }D_{\partial_{\mathbf{\mu}}}%
\mbox{\boldmath{$\gamma$}}%
^{\beta}=-\Gamma_{\mu\alpha}^{\beta}%
\mbox{\boldmath{$\gamma$}}%
^{\alpha}. \label{14}%
\end{equation}
We also write the connection $1$-forms in a coordinate gauge as:%
\begin{equation}
\mathbf{\Gamma}_{\beta}^{\alpha}:=\Gamma_{\mu\beta}^{\alpha}%
\mbox{\boldmath{$\theta$}}%
^{\mu}. \label{15}%
\end{equation}

\subsection{Algebraic Derivatives of Functionals}

Let $X$ $\in\sec%
{\displaystyle\bigwedge\nolimits^{p}}
T^{\ast}M$ . A functional $F$ is a mapping%
\[
F:\sec%
{\displaystyle\bigwedge\nolimits^{p}}
T^{\ast}M\rightarrow\sec%
{\displaystyle\bigwedge\nolimits^{r}}
T^{\ast}M.
\]
When no confusion arises we use a sloppy notation and denote the image
$F(X)\in\sec%
{\displaystyle\bigwedge\nolimits^{r}}
T^{\ast}M$ simply by $F$, or vice versa. Which object we are talking about is
always obvious from the context of the equations where they appear.

Let also $%
\mbox{\boldmath{$\delta$}}%
X\in\sec%
{\displaystyle\bigwedge\nolimits^{p}}
T^{\ast}M$. We define the variation of $F$ as the functional $%
\mbox{\boldmath{$\delta$}}%
F\in\sec%
{\displaystyle\bigwedge\nolimits^{r}}
T^{\ast}M$ given by%
\begin{equation}%
\mbox{\boldmath{$\delta$}}%
F=\lim_{\lambda\rightarrow0}\frac{F(X+\lambda%
\mbox{\boldmath{$\delta$}}%
X)-F(X)}{\lambda}.\label{ad2}%
\end{equation}
Moreover, we define the the algebraic derivative of $F(X)$ relative to $X$\ ,
denoted $\frac{\partial F}{\partial X}$ by:%
\begin{equation}%
\mbox{\boldmath{$\delta$}}%
F=%
\mbox{\boldmath{$\delta$}}%
X\wedge\frac{\partial F}{\partial X}.\label{ad1}%
\end{equation}
Moreover, given $F:\sec%
{\displaystyle\bigwedge\nolimits^{p}}
T^{\ast}M\rightarrow\sec%
{\displaystyle\bigwedge\nolimits^{r}}
T^{\ast}M$, $G:\sec%
{\displaystyle\bigwedge\nolimits^{p}}
T^{\ast}M\rightarrow\sec%
{\displaystyle\bigwedge\nolimits^{s}}
T^{\ast}M$ the variation $%
\mbox{\boldmath{$\delta$}}%
$ satisfies
\begin{equation}%
\mbox{\boldmath{$\delta$}}%
(F\wedge G)=%
\mbox{\boldmath{$\delta$}}%
F\wedge G+F\wedge%
\mbox{\boldmath{$\delta$}}%
G,\label{ad3}%
\end{equation}
and the algebraic derivative satisfies (as it is trivial to verify)
\begin{equation}
\frac{\partial}{\partial X}(F\wedge G)=\frac{\partial F}{\partial X}\wedge
G+(-1)^{rp}F\wedge\frac{\partial G}{\partial X}.\label{ad4}%
\end{equation}

An important property of $%
\mbox{\boldmath{$\delta$}}%
$ is that it commutes with the exterior derivative operator $d$, i.e., for any
given functional $F$%
\begin{equation}
d%
\mbox{\boldmath{$\delta$}}%
F=%
\mbox{\boldmath{$\delta$}}%
dF. \label{ad5}%
\end{equation}

In general we may have functionals depending on several different forms
fields, say, $F(X,Y)\in\sec%
{\displaystyle\bigwedge\nolimits^{r}}
T^{\ast}M$, and $X\in\sec%
{\displaystyle\bigwedge\nolimits^{p}}
T^{\ast}M$, $Y\in\sec%
{\displaystyle\bigwedge\nolimits^{q}}
T^{\ast}M$. In this case we have (using sloop notation):%
\begin{equation}%
\mbox{\boldmath{$\delta$}}%
F=%
\mbox{\boldmath{$\delta$}}%
X\wedge\frac{\partial F}{\partial X}+%
\mbox{\boldmath{$\delta$}}%
Y\wedge\frac{\partial F}{\partial Y}. \label{ad6}%
\end{equation}
An important case happens for a functional $F$ such that $F(X,dX)\in\sec%
{\displaystyle\bigwedge\nolimits^{n}}
T^{\ast}M$ where $n=\dim M$ is the manifold dimension. In this case, for
$U\subset M$, we can write supposing that the variation $%
\mbox{\boldmath{$\delta$}}%
X$ is chosen to be null in the boundary $\partial U$ (or that $\left.
\frac{\partial F}{\partial dX}\right\vert _{\partial U}=0$) and taking into
account Stokes theorem,%
\begin{align}%
\mbox{\boldmath{$\delta$}}%
{\displaystyle\int\nolimits_{U}}
F  &  :=%
{\displaystyle\int\nolimits_{U}}
\mbox{\boldmath{$\delta$}}%
F=%
{\displaystyle\int\nolimits_{U}}
\mbox{\boldmath{$\delta$}}%
X\wedge\frac{\partial F}{\partial X}+%
\mbox{\boldmath{$\delta$}}%
dX\wedge\frac{\partial F}{\partial dX}\nonumber\\
&  =%
{\displaystyle\int\nolimits_{U}}
\mbox{\boldmath{$\delta$}}%
X\wedge\left[  \frac{\partial F}{\partial X}-(-1)^{p}d\left(  \frac{\partial
F}{\partial dX}\right)  \right]  +d\left(
\mbox{\boldmath{$\delta$}}%
X\wedge\frac{\partial F}{\partial dX}\right) \nonumber\\
&  =%
{\displaystyle\int\nolimits_{U}}
\mbox{\boldmath{$\delta$}}%
X\wedge\left[  \frac{\partial F}{\partial X}-(-1)^{p}d\left(  \frac{\partial
F}{\partial dX}\right)  \right]  +%
{\displaystyle\int\nolimits_{\partial U}}
\mbox{\boldmath{$\delta$}}%
X\wedge\frac{\partial F}{\partial dX}\nonumber\\
&  =%
{\displaystyle\int\nolimits_{U}}
\mbox{\boldmath{$\delta$}}%
X\wedge\frac{%
\mbox{\boldmath{$\delta$}}%
F}{%
\mbox{\boldmath{$\delta$}}%
X}, \label{ad7}%
\end{align}
where $\frac{%
\mbox{\boldmath{$\delta$}}%
}{%
\mbox{\boldmath{$\delta$}}%
X}F(X,dX)\in\sec%
{\displaystyle\bigwedge\nolimits^{n-p}}
T^{\ast}M$ is called the \textit{functional derivative} of $F$ and we have:%
\begin{equation}
\frac{%
\mbox{\boldmath{$\delta$}}%
F}{%
\mbox{\boldmath{$\delta$}}%
X}=\frac{\partial F}{\partial X}-(-1)^{p}d\left(  \frac{\partial F}{\partial
dX}\right)  . \label{ad8}%
\end{equation}
When $F=\mathcal{L}$ is a Lagrangian density in field theory $\frac{%
\mbox{\boldmath{$\delta$}}%
\mathcal{L}}{%
\mbox{\boldmath{$\delta$}}%
X}$ is called the Euler -Lagrange functional.\footnote{A detailed theory of
derivatives of non homogeneous multiform functions of multiple non homogeneous
multiform variables may be found in \cite{rodoliv2007}.}

\section{Variation of\textbf{ }the Einstein-Hilbert Lagrangian
Density\ $\mathcal{L}_{EH}$}

We have from $\mathcal{L}_{EH}=\frac{1}{2}\mathcal{R}_{\mathbf{cd}}\wedge
\star(%
\mbox{\boldmath{$\theta$}}%
^{\mathbf{c}}\wedge%
\mbox{\boldmath{$\theta$}}%
^{\mathbf{d}})$,
\begin{align}%
\mbox{\boldmath{$\delta$}}%
\mathfrak{L}_{EH}  &  =\frac{1}{2}%
\mbox{\boldmath{$\delta$}}%
[\mathcal{R}_{\mathbf{cd}}\wedge\star(%
\mbox{\boldmath{$\theta$}}%
^{\mathbf{c}}\wedge%
\mbox{\boldmath{$\theta$}}%
^{\mathbf{d}})]\nonumber\\
&  =\frac{1}{2}%
\mbox{\boldmath{$\delta$}}%
\mathcal{R}_{\mathbf{cd}}\wedge\star(%
\mbox{\boldmath{$\theta$}}%
^{\mathbf{c}}\wedge%
\mbox{\boldmath{$\theta$}}%
^{\mathbf{d}})+\frac{1}{2}\mathcal{R}_{\mathbf{cd}}\wedge%
\mbox{\boldmath{$\delta$}}%
\star(%
\mbox{\boldmath{$\theta$}}%
^{\mathbf{c}}\wedge%
\mbox{\boldmath{$\theta$}}%
^{\mathbf{d}}). \label{v5}%
\end{align}
From Cartan's second structure equation we can write
\begin{align}
&
\mbox{\boldmath{$\delta$}}%
\mathcal{R}_{\mathbf{cd}}\wedge\star(%
\mbox{\boldmath{$\theta$}}%
^{\mathbf{c}}\wedge%
\mbox{\boldmath{$\theta$}}%
^{\mathbf{d}})\nonumber\\
&  =%
\mbox{\boldmath{$\delta$}}%
d\omega_{\mathbf{cd}}\wedge\star(%
\mbox{\boldmath{$\theta$}}%
^{\mathbf{c}}\wedge%
\mbox{\boldmath{$\theta$}}%
^{\mathbf{d}})+%
\mbox{\boldmath{$\delta$}}%
\omega_{\mathbf{ck}}\wedge\omega_{\mathbf{d}}^{\mathbf{k}}\wedge\star(%
\mbox{\boldmath{$\theta$}}%
^{\mathbf{c}}\wedge%
\mbox{\boldmath{$\theta$}}%
^{\mathbf{d}})+\omega_{\mathbf{ck}}\wedge%
\mbox{\boldmath{$\delta$}}%
\omega_{\mathbf{d}}^{\mathbf{k}}\wedge\star(%
\mbox{\boldmath{$\theta$}}%
^{\mathbf{c}}\wedge%
\mbox{\boldmath{$\theta$}}%
^{\mathbf{d}})\nonumber\\
&  =%
\mbox{\boldmath{$\delta$}}%
d\omega_{\mathbf{cd}}\wedge\star(%
\mbox{\boldmath{$\theta$}}%
^{\mathbf{c}}\wedge%
\mbox{\boldmath{$\theta$}}%
^{\mathbf{d}})\label{v6}\\
&  =d[%
\mbox{\boldmath{$\delta$}}%
\omega_{\mathbf{cd}}\wedge\star(%
\mbox{\boldmath{$\theta$}}%
^{\mathbf{c}}\wedge%
\mbox{\boldmath{$\theta$}}%
^{\mathbf{d}})]-%
\mbox{\boldmath{$\delta$}}%
\omega_{\mathbf{cd}}\wedge d[\star(%
\mbox{\boldmath{$\theta$}}%
^{\mathbf{c}}\wedge%
\mbox{\boldmath{$\theta$}}%
^{\mathbf{d}})].\nonumber\\
&  =d[%
\mbox{\boldmath{$\delta$}}%
\omega_{\mathbf{cd}}\wedge\star(%
\mbox{\boldmath{$\theta$}}%
^{\mathbf{c}}\wedge%
\mbox{\boldmath{$\theta$}}%
^{\mathbf{d}})]-%
\mbox{\boldmath{$\delta$}}%
\omega_{\mathbf{cd}}\wedge\lbrack-\omega_{\mathbf{k}}^{\mathbf{c}}\wedge\star(%
\mbox{\boldmath{$\theta$}}%
^{\mathbf{k}}\wedge%
\mbox{\boldmath{$\theta$}}%
^{\mathbf{d}})-\omega_{\mathbf{k}}^{\mathbf{d}}\wedge\star(%
\mbox{\boldmath{$\theta$}}%
^{\mathbf{c}}\wedge%
\mbox{\boldmath{$\theta$}}%
^{\mathbf{k}})]\nonumber\\
&  =d[%
\mbox{\boldmath{$\delta$}}%
\omega_{\mathbf{cd}}\wedge\star(%
\mbox{\boldmath{$\theta$}}%
^{\mathbf{c}}\wedge%
\mbox{\boldmath{$\theta$}}%
^{\mathbf{d}})].\nonumber
\end{align}
Moreover, using the definition of algebraic derivative (Eq.(\ref{ad2})) we
have
\begin{equation}%
\mbox{\boldmath{$\delta$}}%
\star(%
\mbox{\boldmath{$\theta$}}%
^{\mathbf{c}}\wedge%
\mbox{\boldmath{$\theta$}}%
^{\mathbf{d}}):=%
\mbox{\boldmath{$\delta$}}%
\mbox{\boldmath{$\theta$}}%
^{\mathbf{m}}\wedge\frac{\partial\lbrack\star(%
\mbox{\boldmath{$\theta$}}%
^{\mathbf{c}}\wedge%
\mbox{\boldmath{$\theta$}}%
^{\mathbf{d}})}{\partial%
\mbox{\boldmath{$\theta$}}%
^{\mathbf{m}}} \label{v7}%
\end{equation}

Now recalling Eq.(\ref{hodge dual}) of Appendix A we can write
\begin{align}%
\mbox{\boldmath{$\delta$}}%
\star(%
\mbox{\boldmath{$\theta$}}%
^{\mathbf{c}}\wedge%
\mbox{\boldmath{$\theta$}}%
^{\mathbf{d}})  &  =%
\mbox{\boldmath{$\delta$}}%
(\frac{1}{2}\eta^{\mathbf{ck}}\eta^{\mathbf{dl}}\epsilon_{\mathbf{klmn}}%
\mbox{\boldmath{$\theta$}}%
^{\mathbf{m}}\wedge%
\mbox{\boldmath{$\theta$}}%
^{\mathbf{n}})\nonumber\\
&  =%
\mbox{\boldmath{$\delta$}}%
\mbox{\boldmath{$\theta$}}%
^{\mathbf{m}}\wedge(\eta^{\mathbf{ck}}\eta^{\mathbf{dl}}\epsilon
_{\mathbf{klmn}}%
\mbox{\boldmath{$\theta$}}%
^{\mathbf{n}}), \label{v8}%
\end{align}
from where we get%
\begin{equation}
\frac{\partial\star(%
\mbox{\boldmath{$\theta$}}%
^{\mathbf{c}}\wedge%
\mbox{\boldmath{$\theta$}}%
^{\mathbf{d}})}{\partial%
\mbox{\boldmath{$\theta$}}%
^{\mathbf{m}}}=\eta^{\mathbf{ck}}\eta^{\mathbf{dl}}\epsilon_{\mathbf{klmn}}%
\mbox{\boldmath{$\theta$}}%
^{\mathbf{n}}. \label{v9}%
\end{equation}
On the other hand we have recalling Eq.(\ref{7}) of Appendix A%
\begin{align}
&
\mbox{\boldmath{$\theta$}}%
_{\mathbf{m}}\lrcorner\star(%
\mbox{\boldmath{$\theta$}}%
^{\mathbf{c}}\wedge%
\mbox{\boldmath{$\theta$}}%
^{\mathbf{d}})=%
\mbox{\boldmath{$\theta$}}%
_{\mathbf{m}}\lrcorner(\frac{1}{2}\eta^{\mathbf{ck}}\eta^{\mathbf{dl}}%
\epsilon_{\mathbf{klrs}}%
\mbox{\boldmath{$\theta$}}%
^{\mathbf{r}}\wedge%
\mbox{\boldmath{$\theta$}}%
^{\mathbf{s}})\nonumber\\
&  =\eta^{\mathbf{ck}}\eta^{\mathbf{dl}}\epsilon_{\mathbf{klmn}}%
\mbox{\boldmath{$\theta$}}%
^{\mathbf{n}}. \label{v10}%
\end{align}
Moreover, using the fourth formula in Eq.(\ref{440new}) of Appendix A, we can
write
\begin{align}
\frac{\partial\lbrack\star(%
\mbox{\boldmath{$\theta$}}%
^{\mathbf{c}}\wedge%
\mbox{\boldmath{$\theta$}}%
^{\mathbf{d}})}{\partial%
\mbox{\boldmath{$\theta$}}%
^{\mathbf{m}}}  &  =%
\mbox{\boldmath{$\theta$}}%
_{\mathbf{m}}\lrcorner\star(%
\mbox{\boldmath{$\theta$}}%
^{\mathbf{c}}\wedge%
\mbox{\boldmath{$\theta$}}%
^{\mathbf{d}})\nonumber\\
&  =\star\lbrack%
\mbox{\boldmath{$\theta$}}%
_{\mathbf{m}}\wedge(%
\mbox{\boldmath{$\theta$}}%
^{\mathbf{c}}\wedge%
\mbox{\boldmath{$\theta$}}%
^{\mathbf{d}})]=\star(%
\mbox{\boldmath{$\theta$}}%
^{\mathbf{c}}\wedge%
\mbox{\boldmath{$\theta$}}%
^{\mathbf{d}}\wedge%
\mbox{\boldmath{$\theta$}}%
_{\mathbf{m}}). \label{v11}%
\end{align}
Finally,%
\begin{equation}%
\mbox{\boldmath{$\delta$}}%
\star(%
\mbox{\boldmath{$\theta$}}%
^{\mathbf{c}}\wedge%
\mbox{\boldmath{$\theta$}}%
^{\mathbf{d}})=%
\mbox{\boldmath{$\delta$}}%
\mbox{\boldmath{$\theta$}}%
^{\mathbf{m}}\wedge\star(%
\mbox{\boldmath{$\theta$}}%
^{\mathbf{c}}\wedge%
\mbox{\boldmath{$\theta$}}%
^{\mathbf{d}}\wedge%
\mbox{\boldmath{$\theta$}}%
_{\mathbf{m}}). \label{v12}%
\end{equation}
Then using Eq.(\ref{v6}) and Eq.(\ref{v12}) in Eq.(\ref{v5}) we get%
\begin{equation}%
\mbox{\boldmath{$\delta$}}%
\mathfrak{L}_{EH}=\frac{1}{2}d[%
\mbox{\boldmath{$\delta$}}%
\omega_{\mathbf{cd}}\wedge\star(%
\mbox{\boldmath{$\theta$}}%
^{\mathbf{c}}\wedge%
\mbox{\boldmath{$\theta$}}%
^{\mathbf{d}})]+%
\mbox{\boldmath{$\delta$}}%
\mbox{\boldmath{$\theta$}}%
^{\mathbf{m}}\wedge\lbrack\frac{1}{2}\mathcal{R}_{\mathbf{ab}}\wedge\star(%
\mbox{\boldmath{$\theta$}}%
^{\mathbf{a}}\wedge%
\mbox{\boldmath{$\theta$}}%
^{\mathbf{b}}\wedge%
\mbox{\boldmath{$\theta$}}%
_{\mathbf{m}})]. \label{v13}%
\end{equation}

Now,%
\begin{align}
\frac{1}{2}\mathcal{R}_{\mathbf{ab}}\wedge\star(\theta^{\mathbf{a}}%
\wedge\theta^{\mathbf{b}}\wedge\theta_{\mathbf{m}})  &  =-\frac{1}{2}%
\star\lbrack\mathcal{R}_{\mathbf{ab}}\lrcorner(%
\mbox{\boldmath{$\theta$}}%
^{\mathbf{a}}\wedge%
\mbox{\boldmath{$\theta$}}%
^{\mathbf{b}}\wedge%
\mbox{\boldmath{$\theta$}}%
_{\mathbf{m}})]\nonumber\\
&  =-\frac{1}{4}R_{\mathbf{abck}}\star\lbrack(%
\mbox{\boldmath{$\theta$}}%
^{\mathbf{c}}\wedge%
\mbox{\boldmath{$\theta$}}%
^{\mathbf{k}})\lrcorner(%
\mbox{\boldmath{$\theta$}}%
^{\mathbf{a}}\wedge%
\mbox{\boldmath{$\theta$}}%
^{\mathbf{b}}\wedge%
\mbox{\boldmath{$\theta$}}%
_{\mathbf{m}})]\nonumber\\
&  =-\star(\mathcal{R}_{\mathbf{m}}-\frac{1}{2}R%
\mbox{\boldmath{$\theta$}}%
_{\mathbf{m}})=-\star\mathcal{G}_{\mathbf{m}}, \label{v14}%
\end{align}
and so we can write
\begin{equation}
\int%
\mbox{\boldmath{$\delta$}}%
(\mathfrak{L}_{EH}+\mathfrak{L}_{m})=\int%
\mbox{\boldmath{$\delta$}}%
\mbox{\boldmath{$\theta$}}%
^{\mathbf{a}}\wedge(-\star\mathcal{G}_{\mathbf{a}}+\frac{\partial
\mathfrak{L}_{m}}{\partial%
\mbox{\boldmath{$\theta$}}%
^{\mathbf{a}}})=0. \label{v15}%
\end{equation}

\section{Calculation of the Components of $\mathcal{S}_{\lambda}$}

Here, using the powerful Clifford bundle formalism we present two
calculations\footnote{The second one is close to the one given in
\cite{thiwal}.} of the components of $\mathcal{S}_{\lambda}$ given by
Eq.(\ref{22}) in a \textit{coordinate basis} and directly identify Freud's
formula for his quantities $\mathfrak{U}_{\mu}^{\lambda\sigma}$ given in
\cite{freud}.. We start from
\begin{equation}
\star\mathcal{S}_{\lambda}=\frac{1}{2}\mathbf{\Gamma}_{\alpha\beta}\wedge
\star(%
\mbox{\boldmath{$\gamma$}}%
^{\alpha}\wedge%
\mbox{\boldmath{$\gamma$}}%
^{\beta}\wedge%
\mbox{\boldmath{$\gamma$}}%
_{\lambda}). \label{32}%
\end{equation}
Using the third formula in Eq.(\ref{440new}) of Appendix A we can write
\begin{equation}
\star\mathcal{S}_{\lambda}=\mathbf{\Gamma}_{\alpha\beta}\wedge\star(%
\mbox{\boldmath{$\gamma$}}%
^{\alpha}\wedge%
\mbox{\boldmath{$\gamma$}}%
^{\beta}\wedge%
\mbox{\boldmath{$\gamma$}}%
_{\lambda})=\star\left[  \frac{1}{2}\mathbf{\Gamma}_{\alpha\beta}\lrcorner(%
\mbox{\boldmath{$\gamma$}}%
^{\alpha}\wedge%
\mbox{\boldmath{$\gamma$}}%
^{\beta}\wedge%
\mbox{\boldmath{$\gamma$}}%
_{\lambda})\right]
\end{equation}
or
\begin{equation}
\mathcal{S}_{\lambda}=\frac{1}{2}\mathbf{\Gamma}_{\alpha\beta}\lrcorner(%
\mbox{\boldmath{$\gamma$}}%
^{\alpha}\wedge%
\mbox{\boldmath{$\gamma$}}%
^{\beta}\wedge%
\mbox{\boldmath{$\gamma$}}%
_{\lambda}) \label{33}%
\end{equation}
Using now Eq.(\ref{7}) of Appendix A we have%
\begin{equation}
\mathcal{S}_{\lambda}=\frac{1}{2}\left\{  (\mathbf{\Gamma}_{\alpha\beta
}\lrcorner%
\mbox{\boldmath{$\gamma$}}%
^{\alpha})\wedge%
\mbox{\boldmath{$\gamma$}}%
^{\beta}\wedge%
\mbox{\boldmath{$\gamma$}}%
_{\lambda}-(\mathbf{\Gamma}_{\alpha\beta}\lrcorner%
\mbox{\boldmath{$\gamma$}}%
^{\beta})\wedge%
\mbox{\boldmath{$\gamma$}}%
^{\alpha}\wedge%
\mbox{\boldmath{$\gamma$}}%
_{\lambda}+(\mathbf{\Gamma}_{\alpha\beta}\lrcorner%
\mbox{\boldmath{$\gamma$}}%
_{\lambda})\wedge%
\mbox{\boldmath{$\gamma$}}%
^{\alpha}\wedge%
\mbox{\boldmath{$\gamma$}}%
^{\beta}\right\}  \label{34}%
\end{equation}

Now,
\begin{align}
(\mathbf{\Gamma}_{\alpha\beta}\lrcorner%
\mbox{\boldmath{$\gamma$}}%
^{\alpha})\wedge%
\mbox{\boldmath{$\gamma$}}%
^{\beta}\wedge%
\mbox{\boldmath{$\gamma$}}%
_{\lambda}  &  =(%
\mbox{\boldmath{$\gamma$}}%
^{\alpha}\lrcorner\mathbf{\Gamma}_{\alpha\beta})\wedge%
\mbox{\boldmath{$\gamma$}}%
^{\beta}\wedge%
\mbox{\boldmath{$\gamma$}}%
_{\lambda}\nonumber\\
&  \overset{\text{Eq.(\ref{T54})}}{=}%
\mbox{\boldmath{$\gamma$}}%
_{\alpha}\lrcorner(\mathbf{\Gamma}_{\beta}^{\alpha}\wedge%
\mbox{\boldmath{$\gamma$}}%
^{\beta}\wedge%
\mbox{\boldmath{$\gamma$}}%
_{\lambda})+\mathbf{\Gamma}_{\beta}^{\alpha}\wedge(%
\mbox{\boldmath{$\gamma$}}%
_{\alpha}\lrcorner(%
\mbox{\boldmath{$\gamma$}}%
^{\beta}\wedge%
\mbox{\boldmath{$\gamma$}}%
_{\lambda}))\nonumber\\
&  \overset{\text{Eq.(\ref{cartan1})}}{=}-%
\mbox{\boldmath{$\gamma$}}%
_{\alpha}\lrcorner(d%
\mbox{\boldmath{$\gamma$}}%
^{\alpha}\wedge%
\mbox{\boldmath{$\gamma$}}%
_{\lambda})+\mathbf{\Gamma}_{\beta}^{\alpha}\wedge(\delta_{\alpha}^{\beta}%
\mbox{\boldmath{$\gamma$}}%
_{\mu}-g_{\alpha\lambda}%
\mbox{\boldmath{$\gamma$}}%
^{\beta})\nonumber\\
&  \overset{\text{Eq.(\ref{T54})}}{=}-(%
\mbox{\boldmath{$\gamma$}}%
_{\alpha}\lrcorner d%
\mbox{\boldmath{$\gamma$}}%
^{\alpha})\wedge%
\mbox{\boldmath{$\gamma$}}%
_{\lambda}-d%
\mbox{\boldmath{$\gamma$}}%
^{\alpha}\wedge(%
\mbox{\boldmath{$\gamma$}}%
_{\alpha}\lrcorner%
\mbox{\boldmath{$\gamma$}}%
_{\lambda})\nonumber\\
&  +\mathbf{\Gamma}_{\alpha}^{\alpha}\wedge%
\mbox{\boldmath{$\gamma$}}%
_{\mu}-g_{\alpha\lambda}\mathbf{\Gamma}_{\beta}^{\alpha}\wedge%
\mbox{\boldmath{$\gamma$}}%
^{\beta}\nonumber\\
&  =-(%
\mbox{\boldmath{$\gamma$}}%
_{\alpha}\lrcorner d%
\mbox{\boldmath{$\gamma$}}%
^{\alpha})\wedge%
\mbox{\boldmath{$\gamma$}}%
_{\lambda}-g_{\alpha\lambda}(d%
\mbox{\boldmath{$\gamma$}}%
^{\alpha}+\mathbf{\Gamma}_{\beta}^{\alpha}\wedge%
\mbox{\boldmath{$\gamma$}}%
^{\beta})+\mathbf{\Gamma}_{\alpha}^{\alpha}\wedge%
\mbox{\boldmath{$\gamma$}}%
_{\mu}\nonumber\\
&  \overset{\text{Eq.(\ref{cartan1})}}{=}-(%
\mbox{\boldmath{$\gamma$}}%
_{\alpha}\lrcorner d%
\mbox{\boldmath{$\gamma$}}%
^{\alpha})\wedge%
\mbox{\boldmath{$\gamma$}}%
_{\lambda}+\mathbf{\Gamma}_{\alpha}^{\alpha}\wedge%
\mbox{\boldmath{$\gamma$}}%
_{\mu}. \label{35}%
\end{align}
Analogously we find%
\begin{align}
(\mathbf{\Gamma}_{\alpha\beta}\lrcorner%
\mbox{\boldmath{$\gamma$}}%
^{\beta})\wedge%
\mbox{\boldmath{$\gamma$}}%
^{\alpha}\wedge%
\mbox{\boldmath{$\gamma$}}%
_{\lambda}  &  =(%
\mbox{\boldmath{$\gamma$}}%
^{\alpha}\lrcorner d%
\mbox{\boldmath{$\gamma$}}%
_{\alpha})\wedge%
\mbox{\boldmath{$\gamma$}}%
_{\lambda}+\mathbf{\Gamma}_{\alpha}^{\alpha}\wedge%
\mbox{\boldmath{$\gamma$}}%
_{\lambda},\nonumber\\
(\mathbf{\Gamma}_{\alpha\beta}\lrcorner%
\mbox{\boldmath{$\gamma$}}%
_{\lambda})\wedge%
\mbox{\boldmath{$\gamma$}}%
^{\alpha}\wedge%
\mbox{\boldmath{$\gamma$}}%
^{\beta}  &  =(%
\mbox{\boldmath{$\gamma$}}%
_{\lambda}\lrcorner d%
\mbox{\boldmath{$\gamma$}}%
^{\alpha})\wedge%
\mbox{\boldmath{$\gamma$}}%
_{\alpha}-d%
\mbox{\boldmath{$\gamma$}}%
_{\lambda}, \label{36}%
\end{align}
from where we can write%
\begin{equation}
\mathcal{S}_{\mu}=\frac{1}{2}\left[  -(%
\mbox{\boldmath{$\gamma$}}%
_{\alpha}\lrcorner d%
\mbox{\boldmath{$\gamma$}}%
^{\alpha})\wedge%
\mbox{\boldmath{$\gamma$}}%
_{\mu}-(%
\mbox{\boldmath{$\gamma$}}%
^{\alpha}\lrcorner d%
\mbox{\boldmath{$\gamma$}}%
_{\alpha})\wedge%
\mbox{\boldmath{$\gamma$}}%
_{\mu}+(%
\mbox{\boldmath{$\gamma$}}%
_{\mu}\lrcorner d%
\mbox{\boldmath{$\gamma$}}%
^{\alpha})\wedge%
\mbox{\boldmath{$\gamma$}}%
_{\alpha}-d%
\mbox{\boldmath{$\gamma$}}%
_{\mu}\right]  , \label{37}%
\end{equation}
which taking account that $d%
\mbox{\boldmath{$\gamma$}}%
^{\alpha}=d^{2}x^{\alpha}=0$, reduces to%
\begin{equation}
\mathcal{S}_{\mu}=-\frac{1}{2}\left[  (%
\mbox{\boldmath{$\gamma$}}%
^{\alpha}\lrcorner d%
\mbox{\boldmath{$\gamma$}}%
_{\alpha})\wedge%
\mbox{\boldmath{$\gamma$}}%
_{\mu}+d%
\mbox{\boldmath{$\gamma$}}%
_{\mu}\right]  . \label{38}%
\end{equation}

Now from Eq.(\textbf{\ref{13})} valid for a Levi-Civita connection for any
$A\in\sec%
{\displaystyle\bigwedge}
T^{\ast}M\hookrightarrow\mathcal{C\ell(}M,\mathtt{g})$ \ it is
$dA=\mbox{\boldmath$\partial$}\wedge A$. So, we can write (recalling that
$D_{\kappa}g_{\lambda\rho}=0$):%
\begin{align}
d%
\mbox{\boldmath{$\gamma$}}%
_{\mu}  &  =%
\mbox{\boldmath{$\gamma$}}%
^{\kappa}\wedge D_{\partial_{\kappa}}(g_{\mu\rho}%
\mbox{\boldmath{$\gamma$}}%
^{\rho})\nonumber\\
&  =(\partial_{\kappa}g_{\mu\rho}-g_{\mu\beta}\Gamma_{\kappa\rho}^{\beta})%
\mbox{\boldmath{$\gamma$}}%
^{\kappa}\wedge%
\mbox{\boldmath{$\gamma$}}%
^{\rho}\nonumber\\
&  =g_{\beta\rho}\Gamma_{\mu\kappa}^{\beta}%
\mbox{\boldmath{$\gamma$}}%
^{\kappa}\wedge%
\mbox{\boldmath{$\gamma$}}%
^{\rho}\nonumber\\
&  =\delta_{\beta}^{\sigma}\Gamma_{\mu\rho}^{\beta}g^{\kappa\lambda}%
\mbox{\boldmath{$\gamma$}}%
_{\lambda}\wedge%
\mbox{\boldmath{$\gamma$}}%
_{\sigma}\nonumber\\
&  =\frac{1}{2}\left(  \delta_{\beta}^{\sigma}\Gamma_{\mu\rho}^{\beta
}g^{\kappa\lambda}-\delta_{\beta}^{\lambda}\Gamma_{\mu\rho}^{\beta}%
g^{\kappa\sigma}\right)
\mbox{\boldmath{$\gamma$}}%
_{\lambda}\wedge%
\mbox{\boldmath{$\gamma$}}%
_{\sigma}. \label{n1}%
\end{align}

Also,%
\begin{align}%
\mbox{\boldmath{$\gamma$}}%
^{\alpha}\lrcorner d%
\mbox{\boldmath{$\gamma$}}%
_{\alpha}  &  =%
\mbox{\boldmath{$\gamma$}}%
^{\alpha}\lrcorner(g_{\beta\rho}\Gamma_{\alpha\kappa}^{\beta}%
\mbox{\boldmath{$\gamma$}}%
^{\kappa}\wedge%
\mbox{\boldmath{$\gamma$}}%
^{\rho})\nonumber\\
&  =g^{\alpha\kappa}g_{\beta\rho}\Gamma_{\alpha\kappa}^{\beta}%
\mbox{\boldmath{$\gamma$}}%
^{\rho}-g^{\alpha\rho}g_{\beta\rho}\Gamma_{\alpha\kappa}^{\beta}%
\mbox{\boldmath{$\gamma$}}%
^{\kappa}\nonumber\\
&  =(g^{\alpha\kappa}g_{\beta\rho}\Gamma_{\alpha\kappa}^{\beta}-\Gamma
_{\alpha\kappa}^{\alpha})%
\mbox{\boldmath{$\gamma$}}%
^{\rho}, \label{n3}%
\end{align}

and then%
\begin{align}
&  (%
\mbox{\boldmath{$\gamma$}}%
^{\alpha}\lrcorner d%
\mbox{\boldmath{$\gamma$}}%
_{\alpha})\wedge%
\mbox{\boldmath{$\gamma$}}%
_{\mu}\nonumber\\
&  =(\delta_{\mu}^{\sigma}g^{\alpha\kappa}\Gamma_{\alpha\kappa}^{\lambda
}-\delta_{\mu}^{\sigma}g^{\rho\lambda}\Gamma_{\alpha\rho}^{\alpha})%
\mbox{\boldmath{$\gamma$}}%
_{\lambda}\wedge%
\mbox{\boldmath{$\gamma$}}%
_{\sigma}. \label{n4}%
\end{align}
So, we get
\begin{align}
\mathcal{S}_{\mu}  &  =-\frac{1}{2}\left(  \delta_{\beta}^{\sigma}\Gamma
_{\mu\rho}^{\beta}g^{\kappa\lambda}+\delta_{\mu}^{\sigma}g^{\alpha\kappa
}\Gamma_{\alpha\kappa}^{\lambda}-\delta_{\mu}^{\sigma}g^{\rho\lambda}%
\Gamma_{\alpha\rho}^{\alpha}\right)
\mbox{\boldmath{$\gamma$}}%
_{\lambda}\wedge%
\mbox{\boldmath{$\gamma$}}%
_{\sigma}\nonumber\\
&  =\frac{1}{2}\left\{  \frac{1}{2}\det%
\begin{bmatrix}
\delta_{\mu}^{\lambda} & \delta_{\mu}^{\sigma} & \delta_{\mu}^{\iota}\\
g^{\lambda\kappa} & g^{\sigma\kappa} & g^{\iota\kappa}\\
\Gamma_{\kappa\iota}^{\lambda} & \Gamma_{\kappa\iota}^{\sigma} & \Gamma
_{\iota\kappa}^{\iota}%
\end{bmatrix}
\right\}
\mbox{\boldmath{$\gamma$}}%
_{\lambda}\wedge%
\mbox{\boldmath{$\gamma$}}%
_{\sigma}\nonumber\\
&  =\frac{1}{2}\mathcal{S}_{\mu}^{\lambda\sigma}%
\mbox{\boldmath{$\gamma$}}%
_{\lambda}\wedge%
\mbox{\boldmath{$\gamma$}}%
_{\sigma}, \label{n5}%
\end{align}
and then
\begin{equation}
\mathcal{S}_{\mu}^{\lambda\sigma}=\frac{1}{2}\det%
\begin{bmatrix}
\delta_{\mu}^{\lambda} & \delta_{\mu}^{\sigma} & \delta_{\mu}^{\iota}\\
g^{\lambda\kappa} & g^{\sigma\kappa} & g^{\iota\kappa}\\
\Gamma_{\kappa\iota}^{\lambda} & \Gamma_{\kappa\iota}^{\sigma} & \Gamma
_{\iota\kappa}^{\iota}%
\end{bmatrix}
. \label{n6}%
\end{equation}

\subsection{Freud's $\mathfrak{U}_{\mu}^{\lambda\sigma}$}

Now putting%
\begin{equation}
\mathfrak{g}^{\sigma\nu}=\sqrt{-\mathbf{g}}g^{\sigma\nu},\text{ }%
\mathfrak{g}_{\lambda\sigma}=\frac{1}{\sqrt{-\mathbf{g}}}g_{\lambda\sigma}
\label{frackg}%
\end{equation}
we recognize looking at the last formula in Freud's paper \cite{freud} that
his $\mathfrak{U}_{\mu}^{\lambda\sigma}$ is given by
\begin{equation}
\mathfrak{U}_{\mu}^{\lambda\sigma}=\sqrt{-\mathbf{g}}\mathcal{S}_{\mu
}^{\lambda\sigma} \label{FF}%
\end{equation}

\subsection{An Equivalent Formula for Freud's $\mathfrak{U}_{\mu}%
^{\lambda\sigma}$}

We start again our computation of $\mathfrak{U}_{\mu}^{\lambda\sigma}$,
recalling that from (Eq.(\ref{13})) we have for the Hodge coderivative%
\begin{align}
\delta%
\mbox{\boldmath{$\gamma$}}%
^{\alpha}  &  =-\mbox{\boldmath$\partial$}\lrcorner%
\mbox{\boldmath{$\gamma$}}%
^{\alpha}=-%
\mbox{\boldmath{$\gamma$}}%
^{\kappa}\lrcorner(D_{\partial_{\kappa}}%
\mbox{\boldmath{$\gamma$}}%
^{\alpha})\nonumber\\
&  =%
\mbox{\boldmath{$\gamma$}}%
^{\kappa}\lrcorner(\Gamma_{\kappa\rho}^{\alpha}%
\mbox{\boldmath{$\gamma$}}%
^{\rho})=g^{\kappa\rho}\Gamma_{\kappa\rho}^{\alpha}, \label{39}%
\end{align}
and then
\begin{align}%
\mbox{\boldmath{$\gamma$}}%
^{\alpha}\lrcorner d%
\mbox{\boldmath{$\gamma$}}%
_{\alpha}  &  =-2\mathbf{\Gamma}_{\alpha}^{\alpha}+(%
\mbox{\boldmath{$\gamma$}}%
^{\alpha}\lrcorner\mathbf{\Gamma}_{\beta\alpha})%
\mbox{\boldmath{$\gamma$}}%
^{\beta}+%
\mbox{\boldmath{$\gamma$}}%
^{\alpha}\lrcorner\mathbf{\Gamma}_{\alpha\beta})%
\mbox{\boldmath{$\gamma$}}%
^{\beta}\nonumber\\
&  =-2\mathbf{\Gamma}_{\alpha}^{\alpha}+\mathbf{\Gamma}_{\alpha}^{\alpha}+%
\mbox{\boldmath{$\gamma$}}%
_{\alpha}\delta%
\mbox{\boldmath{$\gamma$}}%
^{\alpha}\nonumber\\
&  =-\mathbf{\Gamma}_{\alpha}^{\alpha}+%
\mbox{\boldmath{$\gamma$}}%
_{\alpha}\delta%
\mbox{\boldmath{$\gamma$}}%
^{\alpha}. \label{40}%
\end{align}
Using this result in Eq.(\ref{38}) we get%
\begin{equation}
\mathcal{S}_{\lambda}=-\frac{1}{2}\left(  -\Gamma_{\alpha}^{\alpha}\wedge%
\mbox{\boldmath{$\gamma$}}%
_{\lambda}+(%
\mbox{\boldmath{$\gamma$}}%
_{\alpha}\wedge%
\mbox{\boldmath{$\gamma$}}%
_{\lambda})\delta%
\mbox{\boldmath{$\gamma$}}%
^{\alpha}+d%
\mbox{\boldmath{$\gamma$}}%
_{\lambda}\right)  . \label{41}%
\end{equation}
Recalling that $\mathbf{g}=\det[g_{\alpha\beta}]$ we have the well known
result \cite{landau}%
\begin{equation}
d\mathbf{g}=(\partial_{\alpha}\mathbf{g)}%
\mbox{\boldmath{$\gamma$}}%
^{\alpha}=2\mathbf{g}\Gamma_{\alpha\kappa}^{\kappa}%
\mbox{\boldmath{$\gamma$}}%
^{\alpha}=2\mathbf{g\Gamma}_{\kappa}^{\kappa}, \label{42}%
\end{equation}
and we can write%
\begin{align}
\mathcal{S}_{\lambda}  &  =-\frac{1}{2}\left(  -\frac{d\mathbf{g}}{\mathbf{g}%
}\wedge%
\mbox{\boldmath{$\gamma$}}%
_{\lambda}+(%
\mbox{\boldmath{$\gamma$}}%
_{\alpha}\wedge%
\mbox{\boldmath{$\gamma$}}%
_{\lambda})\delta%
\mbox{\boldmath{$\gamma$}}%
^{\alpha}+d%
\mbox{\boldmath{$\gamma$}}%
_{\lambda}+\frac{1}{2}\frac{d\mathbf{g}}{\mathbf{g}}\wedge%
\mbox{\boldmath{$\gamma$}}%
_{\lambda}\right) \nonumber\\
&  =-\frac{1}{2}\left[  \frac{1}{\mathbf{g}}\left(  -d\mathbf{g}\wedge%
\mbox{\boldmath{$\gamma$}}%
_{\lambda}+\mathbf{g}d%
\mbox{\boldmath{$\gamma$}}%
_{\lambda}+\mathbf{g}(%
\mbox{\boldmath{$\gamma$}}%
_{\alpha}\wedge%
\mbox{\boldmath{$\gamma$}}%
_{\lambda})\delta%
\mbox{\boldmath{$\gamma$}}%
^{\alpha}+\frac{1}{2}d\mathbf{g}\wedge%
\mbox{\boldmath{$\gamma$}}%
_{\lambda}\right)  \right]  . \label{43}%
\end{align}
Now, recalling again that the metric compatibility condition $D_{\kappa
}g_{\lambda\rho}=0$, we have
\begin{align}
&  \frac{1}{2\mathbf{g}}[d\mathbf{g}\wedge%
\mbox{\boldmath{$\gamma$}}%
_{\lambda}+2\mathbf{g}\delta%
\mbox{\boldmath{$\gamma$}}%
^{\alpha}(%
\mbox{\boldmath{$\gamma$}}%
_{\alpha}\wedge%
\mbox{\boldmath{$\gamma$}}%
_{\lambda}]\nonumber\\
&  =\mathbf{\Gamma}_{\kappa}^{\kappa}\wedge%
\mbox{\boldmath{$\gamma$}}%
_{\lambda}+\delta%
\mbox{\boldmath{$\gamma$}}%
^{\alpha}(%
\mbox{\boldmath{$\gamma$}}%
_{\alpha}\wedge%
\mbox{\boldmath{$\gamma$}}%
_{\lambda})\nonumber\\
&  =(\Gamma_{\beta\alpha\kappa}+\Gamma_{\alpha\beta\kappa})g^{\kappa\alpha}%
\mbox{\boldmath{$\gamma$}}%
^{\alpha}\wedge%
\mbox{\boldmath{$\gamma$}}%
_{\lambda}\nonumber\\
&  =(\partial_{\kappa}g_{\alpha\beta})g^{\kappa\alpha}%
\mbox{\boldmath{$\gamma$}}%
^{\alpha}\wedge%
\mbox{\boldmath{$\gamma$}}%
_{\lambda}\nonumber\\
&  =(dg_{\alpha\beta}\lrcorner%
\mbox{\boldmath{$\gamma$}}%
^{\beta})%
\mbox{\boldmath{$\gamma$}}%
^{\alpha}\wedge%
\mbox{\boldmath{$\gamma$}}%
_{\lambda}, \label{44}%
\end{align}
and Eq.(\ref{43}) becomes%
\begin{equation}
\mathcal{S}_{\lambda}=-\frac{1}{2}\left[  \frac{1}{\mathbf{g}}\left(
-d\mathbf{g}\wedge%
\mbox{\boldmath{$\gamma$}}%
_{\lambda}+\mathbf{g}d%
\mbox{\boldmath{$\gamma$}}%
_{\lambda}\right)  +(dg_{\alpha\beta}\lrcorner%
\mbox{\boldmath{$\gamma$}}%
^{\beta})%
\mbox{\boldmath{$\gamma$}}%
^{\alpha}\wedge%
\mbox{\boldmath{$\gamma$}}%
_{\lambda}\right]  . \label{45}%
\end{equation}
However, we also have
\begin{align}
-d\mathbf{g}\wedge%
\mbox{\boldmath{$\gamma$}}%
_{\lambda}+\mathbf{g}d%
\mbox{\boldmath{$\gamma$}}%
_{\lambda}  &  =g_{\lambda\sigma}\mathbf{g}\left[  -\partial_{\beta}%
(\ln\mathbf{g})g^{\nu\beta}g^{\sigma\rho}+g^{\beta\rho}\partial_{\beta
}g^{\sigma\nu}\right]
\mbox{\boldmath{$\gamma$}}%
_{\nu}\wedge%
\mbox{\boldmath{$\gamma$}}%
_{\rho}\nonumber\\
&  =\frac{1}{2}g_{\lambda\sigma}\partial_{\beta}\left[  \mathbf{g}\left(
g^{\sigma\nu}g^{\rho\beta}-g^{\rho\sigma}g^{\nu\beta}\right)  \right]
\mbox{\boldmath{$\gamma$}}%
_{\nu}\wedge%
\mbox{\boldmath{$\gamma$}}%
_{\rho}\nonumber\\
&  -\mathbf{g}(dg_{\alpha\beta}\lrcorner%
\mbox{\boldmath{$\gamma$}}%
^{\beta})%
\mbox{\boldmath{$\gamma$}}%
^{\alpha}\wedge%
\mbox{\boldmath{$\gamma$}}%
_{\lambda}, \label{46'}%
\end{align}
and finally we get
\begin{equation}
\mathcal{S}_{\lambda}=\frac{1}{2}\frac{1}{2(-\mathbf{g)}}g_{\lambda\sigma
}\partial_{\beta}\left[  \mathbf{g}\left(  g^{\sigma\nu}g^{\rho\beta}%
-g^{\rho\sigma}g^{\nu\beta}\right)  \right]
\mbox{\boldmath{$\gamma$}}%
_{\nu}\wedge%
\mbox{\boldmath{$\gamma$}}%
_{\rho}, \label{47}%
\end{equation}
which gives an equivalent expression for the $\mathcal{S}_{\lambda}^{\nu\rho
},$ which is very useful in calculations in GR, e.g., in the calculation of
what is there defined as the \textquotedblleft inertia\textquotedblright%
\ mass\ of a body creating a gravitational field. (see Eq.(\ref{30}) and
below)
\begin{equation}
\mathcal{S}_{\lambda}^{\nu\rho}=\frac{1}{2(-\mathbf{g)}}g_{\lambda\sigma
}\partial_{\beta}\left[  \mathbf{g}\left(  g^{\sigma\nu}g^{\rho\beta}%
-g^{\rho\sigma}g^{\nu\beta}\right)  \right]  . \label{48}%
\end{equation}

From Eq.(\ref{FF}) $\ $above we can then write an equivalent formula for
Freud's $\mathfrak{U}_{\mu}^{\lambda\sigma}$, namely:
\begin{align}
\mathfrak{U}_{\lambda}^{\nu\rho}  &  =\sqrt{-\mathbf{g}}\mathcal{S}_{\lambda
}^{\nu\rho}=\frac{1}{2\sqrt{-\mathbf{g}}}g_{\lambda\sigma}\partial_{\beta
}\left[  \mathbf{g}\left(  g^{\sigma\nu}g^{\rho\beta}-g^{\rho\sigma}%
g^{\nu\beta}\right)  \right] \nonumber\\
&  =-\frac{1}{2}\mathfrak{g}_{\lambda\sigma}\partial_{\beta}\left[  \left(
\mathfrak{g}^{\rho\sigma}\mathfrak{g}^{\nu\beta}-\mathfrak{g}^{\sigma\nu
}\mathfrak{g}^{\rho\beta}\right)  \right]  . \label{49}%
\end{align}

\subsection{The Freud Superpotentials $\mathbf{U}_{\lambda}$}

We also introduce the Freud's superpotentials, i.e., the pseudo $2$-forms
$\mathbf{U}_{\lambda}\in\sec%
{\displaystyle\bigwedge\nolimits^{2}}
T^{\ast}M$, by:
\begin{equation}
\mathbf{U}_{\lambda}=\frac{1}{2}\mathfrak{U}_{\lambda}^{\nu\rho}\in%
\mbox{\boldmath{$\gamma$}}%
_{\nu}\wedge%
\mbox{\boldmath{$\gamma$}}%
_{\rho}. \label{49b}%
\end{equation}

Now, Freud \cite{freud} defined in his Eq.(1)
\begin{equation}
\mathfrak{U}_{\lambda}^{\nu}=\partial_{\rho}\mathfrak{U}_{\lambda}^{\nu\rho
}=-\sqrt{-\mathbf{g}}\Gamma_{\rho\kappa}^{\kappa}\mathcal{S}_{\lambda}%
^{\nu\rho}+\sqrt{-\mathbf{g}}\partial_{\rho}\mathcal{S}_{\lambda}^{\nu\rho}.
\label{50}%
\end{equation}

On the other hand from Eq.(\ref{20}) we have
\begin{equation}
\star^{-1}d\star\mathcal{S}_{\lambda}=\delta\mathcal{S}_{\lambda
}=-\mbox{\boldmath$\partial$}\lrcorner\mathcal{S}_{\lambda}=(-\partial_{\nu
}\mathcal{S}_{\lambda}^{\nu\rho})%
\mbox{\boldmath{$\gamma$}}%
_{\rho}=-\mathcal{G}_{\lambda}-t_{\lambda} \label{51}%
\end{equation}
or%
\begin{equation}
-2\partial_{\kappa}\mathcal{S}_{\nu}^{\kappa\rho}=-2R_{\nu}^{\rho}%
+R\delta_{\nu}^{\rho}-2t_{\nu}^{\rho}. \label{52}%
\end{equation}
Writing
\begin{equation}
\mathfrak{R}_{\nu}^{\rho}=\sqrt{-\mathbf{g}}R_{\nu}^{\rho}\text{,
}\mathfrak{R=}\sqrt{-\mathbf{g}}R\text{, }\mathfrak{t}_{\nu}^{\rho}%
=\sqrt{-\mathbf{g}}t_{\nu}^{\rho} \label{53}%
\end{equation}
and using Eq.(\ref{50}) we have%
\[
-2\sqrt{-\mathbf{g}}\partial_{\kappa}\mathcal{S}_{\nu}^{\rho\kappa}%
+2\sqrt{-\mathbf{g}}\Gamma_{\alpha\kappa}^{\kappa}\mathcal{S}_{\nu}%
^{\rho\alpha}=2\mathfrak{R}_{\nu}^{\rho}-\mathfrak{R}\delta_{\nu}^{\rho
}+2\sqrt{-\mathbf{g}}\Gamma_{\alpha\kappa}^{\kappa}\mathcal{S}_{\nu}%
^{\rho\alpha}+2\mathfrak{t}_{\nu}^{\rho}%
\]
or%
\begin{align}
2\mathfrak{U}_{\nu}^{\rho}  &  =-2\mathfrak{R}_{\nu}^{\rho}+\mathfrak{R}%
\delta_{\nu}^{\rho}-\frac{1}{2}\Gamma_{\kappa\rho}^{\kappa}\mathfrak{g}%
_{\lambda\sigma}\left[  \left(  \mathfrak{g}^{\sigma\nu}\mathfrak{g}%
^{\rho\beta}-\mathfrak{g}^{\rho\sigma}\mathfrak{g}^{\nu\beta}\right)  \right]
,_{\beta}-2\mathfrak{t}_{\nu}^{\rho}\nonumber\\
&  =\delta_{\nu}^{\rho}(\mathfrak{R+L)}-2\mathfrak{R}_{\nu}^{\rho}-\frac{1}%
{2}\Gamma_{\kappa\rho}^{\kappa}\mathfrak{g}_{\lambda\sigma}\left[  \left(
\mathfrak{g}^{\sigma\nu}\mathfrak{g}^{\rho\beta}-\mathfrak{g}^{\rho\sigma
}\mathfrak{g}^{\nu\beta}\right)  \right]  ,_{\beta}-2\mathfrak{t}_{\nu}^{\rho
}-\mathfrak{L}\delta_{\nu}^{\rho}, \label{54}%
\end{align}
which can be written as \cite{freud}%
\begin{equation}
2\mathfrak{U}_{\lambda}^{\nu}=\delta_{\nu}^{\rho}(\mathfrak{R+L)}%
-2\mathfrak{R}_{\nu}^{\rho}+\left(  \Gamma_{\mu\rho}^{\nu}\partial_{\lambda
}\mathfrak{g}^{\mu\rho}-\Gamma_{\kappa\mu}^{\kappa}\partial_{\lambda
}\mathfrak{g}^{\mu\nu}\right)  \label{freud2}%
\end{equation}
with
\begin{equation}
\mathfrak{L}=\mathfrak{g}^{\mu\nu}\left[  \Gamma_{\mu\rho}^{\sigma}%
\Gamma_{\sigma\nu}^{\rho}-\Gamma_{\mu\nu}^{\sigma}\Gamma_{\sigma\rho}^{\rho
}\right]  . \label{lag}%
\end{equation}

\section{The Einstein Energy-Momentum Pseudo $3$-Forms $\star\mathfrak{e}%
^{\lambda}$}

We have from Eq.(\ref{49b})%

\begin{align}
\mbox{\boldmath$\partial$}\lrcorner\mathbf{U}_{\lambda}  &
=\mbox{\boldmath$\partial$}\lrcorner(\sqrt{-\mathbf{g}}\mathcal{S}_{\lambda})=%
\mbox{\boldmath{$\gamma$}}%
^{\kappa}\lrcorner D_{\partial_{\kappa}}(\sqrt{-\mathbf{g}}\mathcal{S}%
_{\lambda})\nonumber\\
&  =-\sqrt{-\mathbf{g}}\Gamma_{\alpha}^{\alpha}\lrcorner\mathcal{S}_{\lambda
}+\sqrt{-\mathbf{g}}\mbox{\boldmath$\partial$}\lrcorner\mathcal{S}_{\lambda
}\nonumber\\
&  =-\sqrt{-\mathbf{g}}\Gamma_{\alpha}^{\alpha}\lrcorner\mathcal{S}_{\lambda
}+\sqrt{-\mathbf{g}}(%
\slT
_{\lambda}+t_{\lambda}), \label{ll1}%
\end{align}

Defining $\mathfrak{T}_{\lambda}$ and $\mathfrak{t}_{\lambda}\in\sec%
{\displaystyle\bigwedge\nolimits^{1}}
T^{\ast}M$ by
\begin{equation}
\mathfrak{T}_{\lambda}=\sqrt{-\mathbf{g}}%
\slT
_{\lambda}, \label{ll1a}%
\end{equation}%
\begin{equation}
\mathfrak{t}_{\lambda}=\sqrt{-\mathbf{g}}(t_{\lambda}-\Gamma_{\alpha}^{\alpha
}\lrcorner\mathcal{S}_{\lambda}) \label{ll1b}%
\end{equation}
or%
\begin{equation}
\star\mathfrak{t}_{\lambda}=\sqrt{-\mathbf{g}}(\star t_{\lambda}%
+\Gamma_{\kappa}^{\kappa}\wedge\star\mathcal{S}_{\lambda}) \label{ll1c}%
\end{equation}
we get
\begin{equation}
\mbox{\boldmath$\partial$}\lrcorner\mathbf{U}_{\lambda}=\mathfrak{T}_{\lambda
}+\mathfrak{t}_{\lambda}. \label{ll4}%
\end{equation}

In components%
\begin{equation}
\partial_{\kappa}\mathfrak{U}_{\lambda}^{\kappa\rho}=\mathfrak{T}_{\lambda
}^{\rho}+\mathfrak{t}_{\lambda}^{\rho} \label{ll4a}%
\end{equation}
Comparing\footnote{Take into account that our definition of the Ricci-tensor
differs by a signal from the one of the quoted author.} Eq.(\ref{ll4a}) with
Eq.(5-5.5) of \cite{trautman} we see that
\begin{equation}
\mathfrak{t}_{\lambda}^{\rho}=\sqrt{-\mathbf{g}}(t_{\lambda}^{\rho}%
-\Gamma_{\alpha\kappa}^{\kappa}\mathcal{S}_{\lambda}^{\alpha\rho})
\label{ll4d}%
\end{equation}
is what is there called the components of the Einstein pseudo-tensor.

Comparing\footnote{See previous footnote.
\par
{}
\par
{}} Eq.(\ref{ll4a}) with Eq.(2.14) of \cite{logunov} we see that what is there
called the components of the Einstein pseudo-tensor are the $\mathfrak{e}%
_{\lambda}^{\rho}$ given by%
\begin{equation}
\mathfrak{e}_{\lambda}^{\rho}=(t_{\lambda}^{\rho}-\Gamma_{\alpha\kappa
}^{\kappa}\mathcal{S}_{\lambda}^{\alpha\rho}). \label{ll4c}%
\end{equation}

Also taking into account Eq.(\ref{19}) we have for the Einstein $3$-forms:
\begin{equation}
\star\mathfrak{e}^{\lambda}=\frac{1}{2}\mathbf{\Gamma}_{\alpha\beta}%
\wedge\lbrack\omega_{\mathbf{\kappa}}^{\mathbf{\lambda}}\wedge\star(%
\mbox{\boldmath{$\theta$}}%
^{\mathbf{\alpha}}\wedge%
\mbox{\boldmath{$\theta$}}%
^{\mathbf{\beta}}\wedge%
\mbox{\boldmath{$\theta$}}%
^{\kappa})+\mathbf{\Gamma}_{\kappa}^{\mathbf{\beta}}\wedge\star(%
\mbox{\boldmath{$\theta$}}%
^{\alpha}\wedge%
\mbox{\boldmath{$\theta$}}%
^{\kappa}\wedge%
\mbox{\boldmath{$\theta$}}%
^{\lambda})+2\mathbf{\Gamma}_{\kappa}^{\kappa}\lrcorner\mathcal{S}^{\lambda}].
\label{LLL}%
\end{equation}

From this we see that Einstein superpotentials are nothing more than the
Freud's superpotentials $\mathbf{U}_{\lambda}$.

\noindent\textbf{Remark 9 \ }The coordinate expression for $\mathfrak{e}%
_{\lambda}^{\rho}$ if you need it can be found in several books, e.g.,
\cite{dirac,logunov}. \ However, important from a historical point of view is
to mention that already in 1917 the famous italian mathematician T.
Levi-Civita\footnote{Yes, the one that gives name to the connection used in
GR.} \ already pointed out \cite{levi} that Einstein solution for the
energy-momentum description of the gravitational field ( the pseudo tensor)
was a nonsequitur.

\subsection{Einstein \textquotedblleft Inertial\textquotedblright\ Mass
$m_{\mathbf{E}}$}

In Section $4$ we defined the "inertial" mass of a body generating a
gravitational field represented by a Lorentzian spacetime with metric $%
\slg
$ by $m_{\mathbf{I}}=$ $\int\star S^{\mu}$, which we comment to be gauge
dependent. Using Eq.(\ref{ll4}) we may define the Landau-Lifshitz inertial
mass by
\begin{equation}
m_{\mathbf{E}}=%
{\displaystyle\int\nolimits_{S^{2}}}
\star\mathbf{U}^{0}. \label{ll7}%
\end{equation}
where $S^{2}$ is a surface of radius $r\rightarrow\infty$. Let us calculate
$\star\mathbf{U}_{\lambda}$ in a coordinate basis. Recalling Eq.(\ref{49}) and
Eq.(\ref{hodge dual}) we have
\begin{align}
\star\mathbf{U}_{\lambda}  &  =-\frac{1}{2}\frac{1}{2\sqrt{-\mathbf{g}}%
}g_{\lambda\sigma}\partial_{\beta}\left[  \mathbf{g}\left(  g^{\rho\sigma
}g^{\nu\beta}-g^{\sigma\nu}g^{\rho\beta}\right)  \right]  \star(%
\mbox{\boldmath{$\gamma$}}%
_{\nu}\wedge%
\mbox{\boldmath{$\gamma$}}%
_{\rho})\nonumber\\
&  =-\frac{1}{2}\frac{1}{2\sqrt{-\mathbf{g}}}g_{\lambda\sigma}g_{\nu\mu
}g_{\rho\varkappa}\partial_{\beta}\left[  \mathbf{g}\left(  g^{\rho\sigma
}g^{\nu\beta}-g^{\sigma\nu}g^{\rho\beta}\right)  \right]  \star(%
\mbox{\boldmath{$\gamma$}}%
^{\mu}\wedge%
\mbox{\boldmath{$\gamma$}}%
^{\varkappa})\nonumber\\
&  =-\frac{1}{2}\frac{1}{2}\frac{\sqrt{-\mathbf{g}}}{2\sqrt{-\mathbf{g}}%
}g_{\lambda\sigma}g_{\nu\mu}g_{\rho\varkappa}\partial_{\beta}\left[
\mathbf{g}\left(  g^{\rho\sigma}g^{\nu\beta}-g^{\sigma\nu}g^{\rho\beta
}\right)  \right]  g^{\mu\varepsilon}g^{\varkappa\tau}\epsilon_{\varepsilon
\tau\alpha\omega}%
\mbox{\boldmath{$\gamma$}}%
^{\alpha}\wedge%
\mbox{\boldmath{$\gamma$}}%
^{\omega}\nonumber\\
&  =-\frac{1}{8}g_{\lambda\sigma}\partial_{\beta}\left[  \mathbf{g}\left(
g^{\rho\sigma}g^{\nu\beta}-g^{\sigma\nu}g^{\rho\beta}\right)  \right]
\epsilon_{\nu\rho\alpha\omega}%
\mbox{\boldmath{$\gamma$}}%
^{\alpha}\wedge%
\mbox{\boldmath{$\gamma$}}%
^{\omega}. \label{ll8}%
\end{align}
Now, for a diagonal metric tensor we have (with $k,m,n=1,2,3$)
\begin{align}
\star\mathbf{U}_{0}  &  =\frac{1}{4}g_{00}\partial_{\beta}\left[
\mathbf{g}\left(  g^{00}g^{\rho\beta}\right)  \right]  \epsilon_{\rho
0\alpha\omega}%
\mbox{\boldmath{$\gamma$}}%
^{\alpha}\wedge%
\mbox{\boldmath{$\gamma$}}%
^{\omega}\label{ll9}\\
&  =-\frac{1}{4}g_{00}\partial_{l}(-\mathbf{g}g^{00}g^{kl})\epsilon_{0kmn}%
\mbox{\boldmath{$\gamma$}}%
^{m}\wedge%
\mbox{\boldmath{$\gamma$}}%
^{n}.\nonumber
\end{align}
\
\begin{equation}
\star\mathbf{U}^{0}=g^{00}\star\mathbf{U}_{0}=-\frac{1}{4}\partial_{l}%
(-g_{11}g_{22}g_{33}g^{kl})\epsilon_{0kmn}%
\mbox{\boldmath{$\gamma$}}%
^{m}\wedge%
\mbox{\boldmath{$\gamma$}}%
^{n}. \label{ll10}%
\end{equation}

Taking into account that if we use \textquotedblleft Cartesian like
coordinates\textquotedblright\ $\{x^{\mu}\}$ (as, e.g., in the
\textit{isotropic} form\footnote{In isotropic Cartesian coordinates the
Schwarzschild solution of the Einstein-Hilbert equation reads (with
$r_{g}=2mG/c^{2}$ in MKS units): $%
\slg
=\left(  \frac{1-r_{g}/4r}{1+r_{g}/4r}\right)  ^{2}dt\otimes dt-(1+r_{g}%
/4r)^{2}%
{\displaystyle\sum\nolimits_{i=1}^{3}}
dx^{i}\otimes dx^{i}.$} of the Schwarzschild solution \cite{weinberg}) we must
define the radial variable of the standard spherical coordinates $($ $r$,
$\theta$, $\varphi)$ by. $r^{2}=-g_{ij}x^{i}x^{j}$.

We parametrize (as it is standard) the surface $S^{2}$ which has equation
$f=x^{i}x_{i}+r^{2}=0$ with the coordinates $(\theta,\varphi)$. The
\ \textquotedblleft Euclidean\textquotedblright\ unitary vector normal to this
surface has thus the components $(n_{1},n_{2},n_{3})$ with $n_{k}=-\frac
{x_{k}}{r}$. Now, we have
\begin{align}
\star\mathbf{U}^{0}  &  =-\frac{1}{4}\partial_{l}(-g_{11}g_{22}g_{33}%
g^{kl})\epsilon_{0kmn}dx^{m}\wedge dx^{n}\nonumber\\
&  =-\frac{1}{2}(U^{1}dx^{2}\wedge dx^{3}+U^{2}dx^{3}\wedge dx^{1}+U^{3}%
dx^{1}\wedge dx^{2}), \label{l11}%
\end{align}
with%
\begin{equation}
U^{k}=\partial_{l}(-g_{11}g_{22}g_{33}g^{kl}). \label{l11b}%
\end{equation}
Since
\begin{equation}
dx^{i}=\frac{\partial x^{i}}{\partial\theta}d\theta+\frac{\partial x^{i}%
}{\partial\varphi}d\varphi\label{l12}%
\end{equation}
we can write Eq.(\ref{l11}) as%
\begin{align}
\star\mathbf{U}^{0}  &  =-\frac{1}{2}\det\left[
\begin{array}
[c]{ccc}%
U^{1} & U^{2} & U^{3}\\
\frac{\partial x^{1}}{\partial\theta} & \frac{\partial x^{2}}{\partial\theta}
& \frac{\partial x^{3}}{\partial\theta}\\
\frac{\partial x^{1}}{\partial\varphi} & \frac{\partial x^{2}}{\partial
\varphi} & \frac{\partial x^{3}}{\partial\varphi}%
\end{array}
\right]  d\theta\wedge d\varphi\nonumber\\
&  =-\frac{1}{2}r^{2}\sin^{2}\theta\det\left[
\begin{array}
[c]{ccc}%
U^{1} & U^{2} & U^{3}\\
\frac{\partial x^{1}}{r\partial\theta} & \frac{\partial x^{2}}{r\partial
\theta} & \frac{\partial x^{3}}{r\partial\theta}\\
\frac{\partial x^{1}}{r\sin^{2}\theta\partial\varphi} & \frac{\partial x^{2}%
}{r\sin^{2}\theta\partial\varphi} & \frac{\partial x^{3}}{r\sin^{2}%
\theta\partial\varphi}%
\end{array}
\right]  d\theta\wedge d\varphi\label{l13}%
\end{align}
Then putting $\vec{U}$ $=(U^{1},U^{2},U^{3})$ and defining moreover the
euclidean orthonormal vectors%
\begin{align}
\vec{e}_{r}  &  =(n_{1},n_{2},n_{3})\nonumber\\
\vec{e}_{\theta}  &  =(\frac{1}{r}\frac{\partial x^{1}}{\partial\theta}%
,\frac{1}{r}\frac{\partial x^{2}}{\partial\theta},\frac{1}{r}\frac{\partial
x^{3}}{\partial\theta}),\label{l14}\\
\text{ }\vec{e}_{\varphi}  &  =(\frac{1}{r\sin^{2}\theta}\frac{\partial x^{1}%
}{\partial\varphi},\frac{1}{r\sin^{2}\theta}\frac{\partial x^{2}}%
{\partial\varphi},\frac{1}{r\sin^{2}\theta}\frac{\partial x^{3}}%
{\partial\varphi}),\nonumber
\end{align}
we can rewrite Eq.(\ref{l13})\ using the standard notation of vector
calculus\footnote{With $\bullet$ denoting the euclidean scalar product and
$\mathbf{\times}$ the vector product.} as:%
\begin{align}
\star\mathbf{U}^{0}  &  =-\frac{1}{2}r^{2}\sin^{2}\theta\text{ }\vec{U}%
\bullet(\vec{e}_{\theta}\text{ }\mathbf{\times}\text{ }\vec{e}_{\varphi
})d\theta\wedge d\varphi\nonumber\\
&  =-\frac{1}{2}r^{2}\sin^{2}\theta\text{ }(\vec{U}\bullet\vec{e}_{r}%
)d\theta\wedge d\varphi\nonumber\\
&  =-\frac{1}{2}r^{2}\sin^{2}\theta U^{i}n_{i}d\theta\wedge d\varphi
\nonumber\\
&  =-\frac{1}{2}\partial_{l}(-g_{11}g_{22}g_{33}g^{li})n_{i}r^{2}\sin
^{2}\theta d\theta\wedge d\varphi. \label{l15}%
\end{align}

Finally, making the radius $r\rightarrow\infty$ we get%
\begin{align}
m_{\mathbf{E}}  &  =%
{\displaystyle\int\nolimits_{S^{2}}}
\star\mathbf{U}^{0}=-\lim_{r\rightarrow\infty}\frac{1}{2}%
{\displaystyle\int\nolimits_{S^{2}}}
\partial_{l}(-g_{11}g_{22}g_{33}g^{lk})n_{k}r^{2}\sin^{2}\theta d\theta\wedge
d\varphi\nonumber\\
&  =-\frac{1}{2}\lim_{r\rightarrow\infty}%
{\displaystyle\int\nolimits_{S^{2}}}
\frac{\partial}{\partial x^{l}}(-g_{11}g_{22}g_{33}g^{kl})n_{k}r^{2}\sin
^{2}\theta d\theta d\varphi\nonumber\\
&  =\frac{1}{2}\lim_{r\rightarrow\infty}%
{\displaystyle\int\nolimits_{S^{2}}}
\frac{x_{k}}{r}\frac{\partial}{\partial x^{l}}(-g_{11}g_{22}g_{33}g^{kl}%
)r^{2}\sin^{2}\theta d\theta d\varphi, \label{ll12}%
\end{align}
a well known result.

For the \textit{isotropic} form of the Schwarzschild metric a simple
calculation shows that $m_{\mathbf{LL}}=m$, the parameter identified as
\textquotedblleft gravitational\textquotedblright\ mass in the solution of
Einstein's equations.

\section{Landau-Lifshitz Energy-Momentum Pseudo $3$-Forms $\star
\mathfrak{l}_{\lambda}$}

Given a coordinate basis associated with a chart with coordinates
$\{x^{\alpha}\}$ covering $U\subset M$ and writing $t^{\mu}=t^{\mu\nu}%
\mbox{\boldmath{$\gamma$}}%
_{\nu}$ given by Eq.(\ref{22}), we immediately discover that the $t^{\mu\nu}$
are not symmetric. So, this object, cannot be used to formulate a
"conservation law " for a \textit{chart} \textit{dependent}\footnote{It is
possible to define global angular momentum $3$-forms only for particular
Lorentzian spacetimes.} angular momentum of matter plus the gravitational
field, i.e., the, $M^{\mu\nu}\in\sec%
{\displaystyle\bigwedge\nolimits^{3}}
T^{\ast}M$%
\begin{equation}
M^{\mu\nu}=x^{\mu}(\star%
\slT
^{\mu}+\star t^{\mu})-x^{\nu}(\star%
\slT
^{\nu}+\star t^{\nu}). \label{angular}%
\end{equation}

In view of this fact let us find an energy-momentum "conservation law"
involving a \textit{symmetric} energy-momentum pseudo tensor.

Define the superpotentials%
\begin{equation}
\mathbf{H}^{\mu}=\mathbf{g}\mathcal{S}^{\mu}=-\sqrt{-\mathbf{g}}%
\mathbf{U}^{\mu}. \label{l1}%
\end{equation}

Then we have
\begin{align}
\mbox{\boldmath$\partial$}\lrcorner(\mathbf{H}^{\mu})  &  =\mathbf{g}%
\mbox{\boldmath$\partial$}\lrcorner\mathcal{S}^{\mu}+2\mathbf{g\Gamma}%
_{\kappa}^{\kappa}\lrcorner S^{\mu}\nonumber\\
&  =(-\mathbf{g)(T}^{\mu}-t^{\mu}-2\mathbf{\Gamma}_{\kappa}^{\kappa}\lrcorner
S^{\mu}\mathbf{)}\label{l2}\\
&  =(-\mathbf{g)(T}^{\mu}-\mathfrak{l}^{\mu}\mathbf{),}%
\end{align}
where
\begin{align}
\star\mathfrak{l}^{\mu}  &  =(\star t^{\mu}+2\mathbf{\Gamma}_{\kappa}^{\kappa
}\lrcorner S^{\mu})\nonumber\\
&  =(\star t^{\mu}-2\mathbf{\Gamma}_{\kappa}^{\kappa}\wedge\star S^{\mu}),
\label{l3}%
\end{align}
are the Landau-Lifshitz energy-momentum $3$-forms as it is obvious comparing
Eq.(\ref{12}) with Eq.(96.15) of \cite{landau}. Also, taking into account
\ Eq.(\ref{22}) we have
\begin{equation}
\star\mathfrak{l}^{\mu}=\frac{1}{2}\mathbf{\Gamma}_{\alpha\beta}\wedge
\lbrack\omega_{\mathbf{\kappa}}^{\mu}\wedge\star(%
\mbox{\boldmath{$\theta$}}%
^{\mathbf{\alpha}}\wedge%
\mbox{\boldmath{$\theta$}}%
^{\mathbf{\beta}}\wedge%
\mbox{\boldmath{$\theta$}}%
^{\kappa})+\mathbf{\Gamma}_{\kappa}^{\mathbf{\beta}}\wedge\star(%
\mbox{\boldmath{$\theta$}}%
^{\alpha}\wedge%
\mbox{\boldmath{$\theta$}}%
^{\kappa}\wedge%
\mbox{\boldmath{$\theta$}}%
^{\mu})+\mathbf{\Gamma}_{\kappa}^{\kappa}\lrcorner\mathcal{S}^{\mu}]
\label{l4}%
\end{equation}

However, the components $\mathfrak{l}^{\mu\nu}$\ are symmetric \cite{landau},
as may be verified by a long calculation.

\subsection{Landau-Lifshitz \textquotedblleft Inertial\textquotedblright\ Mass
$m_{\mathbf{LL}}$}

As a last observation, taking into account Eq.(\ref{ll12}) if we compute
\[
m_{\mathbf{LL}}=%
{\displaystyle\int\nolimits_{S^{2}}}
\star\mathbf{H}^{\mu}%
\]
on the surface of a sphere of radius $r$ and making the radius $r\rightarrow
\infty$ we get for the Schwarzschild solution (in Cartesian isotropic
coordinates) and taking into account that $\lim_{r\rightarrow\infty}%
\sqrt{-\mathbf{g}}=1$,
\begin{equation}
m_{\mathbf{LL}}=\frac{1}{2}\lim_{r\rightarrow\infty}%
{\displaystyle\int\nolimits_{S^{2}}}
\frac{x_{k}}{r}\frac{\partial}{\partial x^{l}}(-g_{11}g_{22}g_{33}g^{kl}%
)r^{2}\sin^{2}\theta d\theta d\varphi=m_{\mathbf{E}}=m
\end{equation}

At this point we end this long Appendix with a comment by Logunov
\cite{logunov}: \smallskip

{\small "it was the fact that "inertial" mass coincides with gravitational
mass that gave grounds for asserting that they are equal in GR, to".
\smallskip}

Indeed, in their celebrated textbook, Landau and Lifshitz \cite{landau} wrote
at page 334:\smallskip

{\small \textquotedblleft...}$P^{0}=m${\small , a result\footnote{{\small In
\cite{landau} $m=Mc$.}} which was naturally to be expected. It is an
expression of the equality of \textquotedblleft
gravitational\textquotedblright\ and \textquotedblleft
inertial\textquotedblright\ mass ( \textquotedblleft
gravitational\textquotedblright\ mass is the mass that determine the
gravitational field produced by the body, the same mass that appears in the
metric tensor of the gravitational field, or in particular, in Newton's law;
\textquotedblleft inertial" mass is the mass that determines the ratio of
energy momentum of the body; in particular, the rest energy of the body is
equal to the mass multiplied by }$c^{2}$.{\small "\smallskip}

However, as discussed in \cite{boro, logunov} the $%
{\displaystyle\int\nolimits_{S^{2}}}
\star\mathbf{H}^{0}$ (or $%
{\displaystyle\int\nolimits_{S^{2}}}
\star\mathbf{U}^{0}$) being the integral of a gauge dependent quantity depends
on the coordinate chart chosen for its computation, and we can easily build
examples in which the \textquotedblleft inertial\textquotedblright\ mass is
\textit{different} from the \textquotedblleft gravitational\textquotedblright%
\ mass, violating the main Einstein's heuristic guide to GR, namely the
equality of both masses. This results makes one to understand the reason of
Sachs\&Wu statement quoted above.\medskip

\noindent\textbf{Acknowledgement} \ The authors will be grateful to any one
that inform them of any misprints or eventual errors.

\end{document}